\documentclass[%
 reprint,
 amsmath,amssymb,
 aps,
prb,
]{revtex4-2}

\usepackage{xcolor}
\usepackage{physics}
\usepackage{graphicx}% Include figure files
\usepackage{dcolumn}% Align table columns on decimal point
\usepackage{bm}% bold math

\usepackage[colorlinks=true,allcolors=blue]{hyperref}
\usepackage{booktabs}
\usepackage{tabularray}
\usepackage{enumitem}
\usepackage{physics}
\usepackage{soul}
\usepackage{xfrac}
\DeclareMathAlphabet\mathbfcal{OMS}{cmsy}{b}{n}

\begin{document}

\preprint{APS/123-QED}

\title{Spin-correlation dynamics:\\ A semiclassical framework for nonlinear quantum magnetism}
 
%\title{Theory of spin correlations for ultrafast nonlinear quantum magnetism}% Force line breaks with \\

\author{Lukas Körber}\email{lukas.koerber@ru.nl}

\affiliation{Radboud University, Institute of Molecules and Materials, Heyendaalseweg 135, 6525 AJ Nijmegen, The Netherlands}

\author{Pim Coenders}%
 
\affiliation{Radboud University, Institute of Molecules and Materials, Heyendaalseweg 135, 6525 AJ Nijmegen, The Netherlands}


\author{Johan H. Mentink}\email{johan.mentink@ru.nl}
\affiliation{Radboud University, Institute of Molecules and Materials, Heyendaalseweg 135, 6525 AJ Nijmegen, The Netherlands}


\date{\today}% It is always \today, today,

\begin{abstract}

Classical nonlinear theories are highly successful in describing far-from-equilibrium dynamics of magnets, encompassing phenomena such as parametric resonance, ultrafast switching, and even chaos. However, at ultrashort length and time scales, where quantum correlations become significant, these models inevitably break down. While numerous methods exist to simulate quantum many-body spin systems, they are often limited to near-equilibrium conditions, capture only short-time dynamics, or obscure the intuitive connection between nonlinear behavior and its geometric origin in the $\mathfrak{su}(2)$ spin algebra. To advance nonlinear magnetism into the quantum regime, we develop a theory in which semiclassical spin correlations, rather than individual spins, serve as the fundamental dynamical variables. Defined on the bonds of a bipartite lattice, these correlations are inherently nonlocal, with dynamics following through a semiclassical mapping that preserves the original spin algebra. The resulting semiclassical theory captures nonlinear dynamics that are entirely nonclassical and naturally accommodates phenomenological damping at the level of correlations, which is typically challenging to include in quantum methods. As an application, we focus on Heisenberg antiferromagnets, which feature significant quantum effects. We predict nonlinear scaling of the mean frequency of quantum oscillations in the Néel state with the spin quantum number $S$. These have no classical analog and exhibit features reminiscent of nonlinear parametric resonance, fully confirmed by exact diagonalization. The predicted dynamical features are embedded in the geometric structure of the semiclassical phase space of spin correlations, making their physical origin much more transparent than in full quantum methods. With this, semiclassical spin-correlation dynamics provide a foundation for exploring nonlinear quantum magnetism.

%\begin{description}
%\end{description}
\end{abstract}

\maketitle

\section{Introduction} 
\label{sec:intro}

Although magnetism is fundamentally of quantum origin, classical nonlinear models successfully capture a vast range of static and dynamic phenomena in magnetically ordered systems. Prominent examples include spin waves \cite{gurevichMagnetizationOscillationsWaves2000}, topological solitons \cite{kosevichMagneticSolitons1990} and domains \cite{hubertMagneticDomainsAnalysis1998}, as well as strongly nonlinear effects such as magnetization reversal \cite{backMinimumFieldStrength1999,raduTransientFerromagneticlikeState2011,hertelUltrafastNanomagneticToggle2007}, parametric resonance \cite{bloembergenRelaxationEffectsFerromagnetic1952,suhlTheoryFerromagneticResonance1957}, self-oscillations \cite{slavinNonlinearAutoOscillatorTheory2009,chenSpinTorqueSpinHallNanoOscillators2016}, and even chaotic dynamics in driven magnets \cite{deaguiarObservationSubharmonicRoutes1986,williameChaoticDynamicsMacrospin2019}. The intrinsic nonlinearity of magnetism originates from the fact that, in quantum theory, spin operators form the $\mathfrak{su}(2)$ algebra of angular momenta, the generators of rotations in space. Classical theories such as atomistic spin dynamics \cite{AtomisticSpinDynamics2017} and micromagnetism \cite{brownjr.Micromagnetics1963} embody this property by representing spins or magnetization as axial vectors of fixed length that evolve according to torque equations. In doing so, they provide an intuitive, geometrically transparent picture of nonlinear spin dynamics in terms of precession and damping, influenced by effective magnetic fields.

Nevertheless, as one probes ever shorter length and time scales, set ultimately by the exchange interactions between microscopic spins in the crystal lattice, classical theories are expected to break down and quantum effects such as zero-point fluctuations become increasingly relevant. Recent experimental advances have made quantum spin dynamics directly accessible, unveiling a range of phenomena in quantum magnonics \cite{jiangIntegratingMagnonsQuantum2023}, 
where spin waves act as transducers in quantum circuits \cite{wangQuantumInterfacesNanoscale2021,bejaranoParametricMagnonTransduction2024}, or serve as witnesses of squeezing and entanglement \cite{zhaoMagnonSqueezingAntiferromagnet2004,zhaoMagnonSqueezingAntiferromagnetic2006,kamraAntiferromagneticMagnonsHighly2019,fabianiUltrafastDynamicsEntanglement2022,azimi-mousolouMagnonShakeupEntanglement2025}. Furthermore, direct optical manipulation of exchange interactions can induce the dynamics of entangled magnon pairs that classical spin models cannot capture \cite{bossiniLaserdrivenQuantumMagnonics2019,formisanoCoherentTHzSpin2024a}, exhibiting non-perturbative magnon-magnon interactions that are not yet fully understood  \cite{fabianiSupermagnonicPropagationTwoDimensional2021,boumanTimedependentSchwingerBoson}. It remains unclear how to extend classical spin theories, which are so successful in describing nonlinear dynamics, to include intrinsic dynamical quantum effects.

In principle, the quantum dynamics of magnetically ordered media can be described using many-body quantum theory. For example, perturbative techniques that describe spin dynamics through magnon-magnon interactions via bosonization can capture quantum effects but are limited in modeling nonlinear dynamics far from equilibrium. Semiclassical techniques, such as the truncated Wigner transform and its extensions \cite{schachenmayerManyBodyQuantumSpin2015,minkHybridDiscretecontinuousTruncated2022,nagaoTwodimensionalCorrelationPropagation2025}, can capture important quantum features but ultimately rely on Monte Carlo sampling in classical phase space. Fully quantum simulation methods, including quantum Monte-Carlo techniques \cite{foulkesQuantumMonteCarlo2001}, tensor-network methods \cite{orusPracticalIntroductionTensor2014} or neural quantum states \cite{carleoSolvingQuantumManybody2017}, instead evolve a sampled or compressed quantum state vector under the Schrödinger equation. In both cases, the dynamical behavior of spin observables -- that is, the dynamics of their semiclassical or quantum expectation values -- has to be extracted \textit{a posteriori}. As a consequence, the direct connection between spin dynamics and the underlying geometrical structure of the $\mathfrak{su}(2)$ algebra responsible for the plethora of nonlinear phenomena in magnetism is obscured. This makes it difficult to connect the intuition developed in classical nonlinear spin dynamics to its quantum counterpart. It is therefore desirable to formulate a description of nonlinear magnetism that incorporates nonclassical effects directly at the level of (semiclassical) observables.

The primary focus of this paper is to develop a new approach to quantum spin dynamics in magnetically ordered systems that directly incorporates the geometrical structure of $\mathfrak{su}(2)$ at the level of semiclassical observables. Instead of relying on a systematic approximation of the quantum many-body problem, we develop a theory of nonlinear magnetism that incorporates nonclassical dynamics in terms of spin observables that can be interpreted (semi)classicaly. A crucial conceptual stepping stone in this new formulation is that we focus on two-spin rather than single-spin observables. Hence, instead of relying on classical spin vectors, we propose a semiclassical theory in which the dynamical variables are spin correlations defined on the bonds of a bipartite lattice. We show that within our theory, the correlations, along with the bond-wise antiferromagnetic Neel vector and magnetization components, form a complete set of variables that dictate the dynamics of magnets with localized spins beyond what can be modeled with classical spins. 

We demonstrate the capabilitiy to capture dynamics that are both nonclassical and nonlinear at the example of Heisenberg antiferromagnets which are known to exhibit significant quantum effects. Confirmed by exact diagonalization, we show that the quantum oscillations starting from the antiferromagnetic Néel state exhibit a nonlinear scaling of their mean frequencies in the value of the spin $S$. Moreover, the dynamics feature frequency doublings reminescent to the nonlinear parametric resonance known from classical spins, where one component oscillates with twice the frequency. These features are neither captured by classical magnetism nor by quantum linear spin-wave theory. The predicted dynamics are directly engrained in the geometrical structure of the phase space of semiclassical spin correlations and are, therefore, much more transparent as compared to full quantum methods. Furthermore, due to its geometric similarity with classical magnetism, our semiclassical theory naturally incorporates dissipative terms analogous to the damping term in the Landau–Lifshitz formulation. This enables us to describe relaxation and equilibrium directly at the level of spin correlations, something which is typically more involved in full quantum methods based on Lindblad or density-matrix dynamics.

Our approach is inspired by, yet extends beyond, earlier formulations of quantum spin dynamics that treat spin correlations as the fundamental dynamical variables \cite{fedianinSelectionRulesUltrafast2023,formisanoCoherentTHzSpin2024a}. Those works describe linear magnon-pair excitations governed by an emergent $\mathfrak{su}(1,1)$ algebra derived from two-boson Perelomov operators, which constitute an approximation to the full nonlinear quantum spin dynamics \cite{perelomovCoherentStatesSymmetric1975,bossiniLaserdrivenQuantumMagnonics2019}. In contrast, our semiclassical framework retains the complete nonlinearity encoded in the underlying $\mathfrak{su}(2)$ algebra. By reformulating spin dynamics on the level of correlations, our theory provides a stepping stone towards a better understanding of the dynamics of quantum spin systems far from equilibrium. 

The paper is organized as follows. In Sec.~\ref{sec:roadmap}, we review the conventional classical limit of magnetism to motivate our strategy for constructing a semiclassical theory of spin correlations. After fixing notation in Sec.~\ref{sec:notation}, we introduce the bond correlation operators and their commutation relations in Sec.~\ref{sec:neel-operators}, before obtaining their semiclassical equations of motion in Sec.~\ref{sec:eom-semiclassical}. This leads to a quantum-classical correspondence for correlations that we use to derive dissipative terms in their equations of motion in Sec.~\ref{sec:damping}. We conclude the derivation of the theory in Sec.~\ref{sec:full-correlation} by providing a prescription to calculate the full spin correlation. Finally, we apply our theory to Heisenberg antiferromagnets by studying their semiclassical ground states in Sec.~\ref{sec:ground-states} and non-equilibrium dynamics in Sec.~\ref{sec:neel-oscillations}, before concluding an outlook in Sec.~\ref{sec:conclusion_outlook}. 

\section{Strategy}\label{sec:roadmap}

To motivate the derivation of our spin-correlation theory, we briefly review the classical dynamics of a spin system derived from a quantum theory. Given a spin Hamiltonian $\hat{H}$, the quantum Landau-Lifshitz equation describing the dynamics of a single spin operator $\hat{\boldsymbol{S}}$ in the Heisenberg picture is straightforwardly obtained from the Heisenberg equation
\begin{equation}\label{eq:quantum-ll}
    \dv{}{t}\hat{\boldsymbol{S}}_i = \frac{1}{i\hbar}\big[\hat{\boldsymbol{S}}_i,\hat{H}\big] = \hat{\boldsymbol{S}}_i \times \pdv{\hat{H}}{\hat{\boldsymbol{S}}_i}
\end{equation}
by using the well-known commutation relations
\begin{equation}\label{eq:su2-algebra}
    \big[\hat{S}^\alpha_i,\hat{S}^\beta_j\big] = i \hbar\, \delta_{ij} \epsilon_{\alpha\beta\gamma}\, \hat{S}^\gamma_i \qquad \alpha,\beta,\gamma =x,y,z
\end{equation}
of the $\mathfrak{su}(2)$ Lie algebra for two spins at sites $i$ and $j$. Here $\delta_{ij}$ is the Kronecker delta and $\epsilon_{\alpha\beta\gamma}$ the Levi-Civita symbol. The classical limit of Eq.~\eqref{eq:quantum-ll} can be carried out by taking the limit $\hbar \rightarrow 0$, while keeping the product {$\mathcal{S} \equiv S\hbar$} constant. In this way, one obtains the well-known {mapping from a quantum to a classical theory}
\begin{subequations}\label{eq:classical-limit}
    \begin{align}
    \hat{\boldsymbol{S}}_i, \hat{H}  \ & \mapsto\ \mathbfcal{S}_i, \mathcal{H}_\mathrm{cl}\\
    \frac{1}{i\hbar}\big[\hat{\boldsymbol{S}}_i,\hat{H}\big] \ & \mapsto\ \qty{{\mathbfcal{S}}_i,\mathcal{H}_\mathrm{cl}},\\
    S(S+1) \ & \mapsto \ S^2,\label{eq:replace-spin-classical}
    \end{align}
\end{subequations}
through which the commutator is replaced with the Poisson bracket $\{\cdot,\cdot\}$ acting on the observables $\mathbfcal{S}_i$ and $\mathcal{H}_\mathrm{cl}$ defined on the corresponding classical phase space. For $n$ spins, the underlying set of this classical phase space is the product of $n$ spheres with radius {$\abs{\mathbfcal{S}_i}=\mathcal{S}$} (for simplicity, we assume homogeneous spin length). For any two functions $\mathcal{F}$ and $\mathcal{G}$ on this classical phase space \cite{riosPoissonAlgebraClassical2014}, the Poisson bracket is given as
\begin{equation}\label{eq:poissonbracket}
    \qty{\mathcal{F},\mathcal{G}} = - \sum\limits_{i=1}^n \mathbfcal{{S}}_i \cdot\qty(\pdv{\mathcal{F}}{\mathbfcal{{S}}_i}\times \pdv{\mathcal{G}}{\mathbfcal{{S}}_i}),
\end{equation}
which is equivalent to the classical Poisson algebra $\{\mathcal{S}_i^\alpha,\mathcal{S}_j^\beta\} = \delta_{ij}\epsilon_{\alpha\beta\gamma}\mathcal{S}_i^\gamma$, and analogous to the commutation relations Eq.~\eqref{eq:su2-algebra} that describe the local rotational symmetry in spin space. It is important to note that the nonlinearity in magnetism originates exactly from this symmetry, \textit{i.e.} that the phase space of each spin is a sphere, not a plane. Note that various semiclassical treatments refrain from mapping the spin value according to Eq.~\eqref{eq:replace-spin-classical} \cite{kleinNoteClassicalSpinWave1950,andersonApproximateQuantumTheory1952}, allowing to capture corrections to groundstate energies due to zero-point fluctuations. This, however, does not induce any qualitative changes to the dynamics of classical single-spin observables as their Poisson brackets remain invariant with respect to this change.

Despite its effectiveness in describing nonlinear spin dynamics, the classical equation of motion is fundamentally incapable of capturing entanglement effects.
{To illustrate this, consider the spin dynamics governed by Eq.~\eqref{eq:quantum-ll} in the Heisenberg picture for some entangled state $\ket{\psi}$. Entanglement strictly implies that the expectation value of the right-hand side of Eq.~\eqref{eq:quantum-ll} cannot be factorized into products, 
\begin{equation}\label{eq:decoupling-spins}
\left\langle\psi\middle\vert{\hat{\boldsymbol{S}}_i \times \pdv{\hat{H}}{\hat{\boldsymbol{S}}_i}}\middle\vert\psi\right\rangle
\neq
\expval{\hat{\boldsymbol{S}}_i}{\psi} \times
\left\langle\psi\middle\vert{\pdv{\hat{H}}{\hat{\boldsymbol{S}}_i}}\middle\vert\psi\right\rangle.
\end{equation}
Note that the converse is not generally true. Such a factorization is required by a fully classical description to obtain closed equations of motion for the expectation values $\langle \hat{\boldsymbol{S}}_i \rangle \approx \mathbfcal{S}_i$. Consequently, classical magnetism cannot describe the dynamics of spin observables in entangled states.

In this work, rather than performing the quantum-to-classical mapping at the level of individual spins, we adopt an alternative route and carry out such a mapping at the level of spin correlations. A key ingredient of this approach is to go beyond single-particle operators and instead focus on two-particle operators. Analogous to classical magnetism, this approach yields a closed set of equations of motion, but recovers nonclassical dynamics beyond the mapping based on single-spin observables.

\section{Theory}

Building on the strategy inspired by the purely classical limit of magnetism, we now develop a semiclassical theory of spin correlations. The first step is to introduce a set of spin-correlation operators defined on the bonds of a bipartite lattice and to determine their mutual commutation relations. Unlike spin operators on different sites, these bond operators generally do not commute, which leads to the generation of additional {correlation operators} in the resulting Heisenberg equations of motion. To solve these equations, we apply a {semiclassical mapping} that yields Hamiltonian equations for {bond-wise} spin correlations{, expressed in terms of Poisson brackets}. A key advantage of this formulation is that it naturally accommodates the introduction of phenomenological damping terms -- analogous to the Gilbert damping in classical magnetism -- directly on the level of correlation dynamics. As a result, we are able to determine semiclassical equilibrium states in terms of spin correlations.

\subsection{Notation}\label{sec:notation}

Throughout the paper, we use a collection of mathematical notation and special symbols which we summarize here briefly for the readers convenience.

\begin{description}[labelwidth=1.5cm, leftmargin=!]
    \item[$\hat{O}, \hat{\boldsymbol{O}}$] Quantum mechanical operators and vector operators.
    \item[$\mathcal{O}, {\mathbfcal{O}}$] Scalar- or vector functions on phase space.
    \item[$\expval{...}$] Quantum expectation value $\expval{...}{\psi}$ in some state $\ket{\psi}$.
    \item[$\lbrack \hat{Q},\hat{P}\rbrack_{(+)}$]     (Anti)commutator of two operators $\hat{Q}$ and $\hat{P}$.  
    \item[$\{\mathcal{Q},\mathcal{P}\}_{(+)}$] Poisson bracket (dissipative bracket) of two phase-space functions $\mathcal{Q}$ and $\mathcal{P}$ [see definitions in Eqs.~\eqref{eq:poissonbracket} and \eqref{eq:dissipative-bracket}].
    \item[$\mathbb{A}, \mathbb{B}$] The two sublattices of a bipartite lattice.
    \item[${i,j,k}$] Latin subscript indices of individual lattice sites, as in $\hat{S}^z_i$.
    \item[${\nu,\mu}$] Greek subscript indices of undirected bonds connecting two sites $i$ and $j$, as in $\hat{N}^z_\nu$. Undirected here means that we indentify $(i,j)$ and $(j,i)$ as the same bond [see defitions in Sec.~\ref{sec:bond-notation}].
    \item[${\alpha,\beta,\gamma}$] Greek superscript component of vector operator/function, as in $\hat{S}^\alpha$ (\textit{here}, $\alpha=x,y,z$).
    \item[$\hat{S}^\alpha_i, \mathcal{S}^\alpha_i$] Component $\alpha$ of the quantum spin $\hat{\boldsymbol{S}}_i$ or classical spin  $\mathbfcal{S}_i$ at lattice site $i$. 
    \item[$\hat{N}^\alpha_\nu, \mathcal{N}^\alpha_\nu$] Component $\alpha$ of the Néel operators or semiclassical spin correlations defined on a bond $\nu$, as in $\hat{N}^\pm_\nu$ [see definitions in Sec.~\ref{sec:neel-operators}]. As a special case, the fourth component is denoted as $\mathcal{M}^z_\nu$ ($z$-component of bond magnetization).
\end{description}

\subsection{Spin-spin correlation operators}\label{sec:neel-operators}

In this section, we introduce the Néel correlation operators $\hat{N}^\alpha_\nu$, which are the central objects in our theory. In contrast to spins, these correlation operators are defined on the \textit{bonds} $\nu = (i,j)$ of a bipartite lattice with homogeneous spin $S$. Our formulation is adapted to Hamiltonians with nearest-neighbor (inter-sublattice) interactions, but its generalization does not require any conceptual differences. Before defining the bond-index notation any further, we focus on the simplest case of correlation operators on an isolated bond (dimer) and derive the commutation relations necessary for the equations of motion. 

\subsubsection{Case of a single bond}

The inner product of two spins at site $i$ and $j$ can be written with the spin-ladder operators $\hat{S}^\pm_{i} = \hat{S}^x_{i} \pm i\hat{S}^y_{i}$ as
\begin{equation}\label{eq:spindot}
    \hat{\boldsymbol{S}}_i\cdot \hat{\boldsymbol{S}}_j = \tfrac{1}{2}\Big({\hat{S}^+_i\hat{S}^-_j} + {\hat{S}^-_i\hat{S}^+_j}\Big) + \hat{S}^z_i\hat{S}^z_j,
\end{equation}
Within the representation Eq.~\eqref{eq:spindot}, we suggestively define the \textit{Néel ladder operators} as
\begin{subequations}\label{eq:neel-operators}
\begin{equation}
    \hat{N}^\pm = \frac{\hat{S}^\pm_{i}\hat{S}^\mp_j}{2},
\end{equation}
where we have surpressed the bond index of the individual dimer for notational efficiency. 
These operators $\hat{N}^\pm$ raise (lower) the spin on one site while simultaneously lowering (raising) its neighbor. Thus, they represent a correlated two-spin process, in which the eigenvalue of the $z$-projection difference $\hat{S}_i^z - \hat{S}_j^z$ changes. In addition, we introduce the \textit{bond magnetization} and \textit{bond Néel vector} components as 
\begin{align}
    \hat{M}^z & = \tfrac{1}{2}\big(\hat{S}^z_{i} + \hat{S}^z_{j}\big)\\
    \hat{N}^z & = \tfrac{1}{2}\big(\hat{S}^z_{i} - \hat{S}^z_{j}\big).
\end{align}
\end{subequations}
In terms of these two operators, the longitudinal part $\hat{S}^z_i\hat{S}^z_j$ of the spin product can be decomposed as 
\begin{equation}
    \hat{S}^z_i\hat{S}^z_j = \big(\hat{M}^z\big)^2 - \big(\hat{N}^z\big)^2
\end{equation}
which allows us to write Eq.~\eqref{eq:spindot} fully in terms of the Néel operators. In order to predict their dynamcis, next, we derive the exact commutation relations of the two-spin correlation operators. These relations form the foundation for the fully nonlinear equations of motion developed below. Since the correlation operators are built from single-spin operators, their algebra follows directly from the underlying 
$\mathfrak{su}(2)$ structure. We find}
\begin{subequations}\label{eq:single-bond-commutators}
    \begin{align}
    \big[\hat{N}^z,\hat{M}^z\big] & = 0 \\
    \big[\hat{M}^z,\hat{N}^\pm\big]& = 0 \\
    \big[\hat{N}^z,\hat{N}^\pm\big] & =  \pm \hbar \hat{N}^\pm \label{eq:ladder-commutator-single}  \\
    \big[\hat{N}^+,\hat{N}^-\big] &= \hbar \hat{N}^z\Big[S(S+1)\hbar^2\hat{I} \\ & \qquad \qquad +(\hat{M}^z)^2-(\hat{N}^z)^2\Big]\notag 
\end{align}
\end{subequations}
where we have also used that $\hat{\boldsymbol{S}}^2_i = S(S+1)\hbar^2\hat{I}$. Equation~\eqref{eq:ladder-commutator-single} shows that the operators $\hat{N}^\pm$ indeed act as ladder operators on the antiferromagnetic Néel vector. When linearized around a Néel state (see Appendix~\ref{sec:su11}), the commutation relations Eq.~\eqref{eq:single-bond-commutators} reduce to an  $\mathfrak{su}(1,1)$ algebra of Perelomov operators that were previously associated with creation and annihilation of entangled magnon pairs \cite{perelomovCoherentStatesSymmetric1975,bossiniLaserdrivenQuantumMagnonics2019}, illustrating how our theory is a generalization of previous work. Without linearization,{the commutation relations of the Néel operators} allow us to derive a set of Heisenberg equations for the full correlation dynamics on a single bond. For example, for a two-spin Heisenberg Hamiltonian
\begin{equation}
\begin{split}
        \hat{H} & = J\hat{\boldsymbol{S}}_1\cdot \hat{\boldsymbol{S}}_2 \\ & = J\qty[\hat{N}^+ + \hat{N}^- + \big(\hat{M}^z\big)^2 - \big(\hat{N}^z\big)^2]
\end{split}
\end{equation}
with exchange constant $J$
we obtain the Heisenberg equations
\begin{equation}\label{eq:heisenbergspincorrgeneral}
    \dv{{\hat{N}^\alpha}}{t} = \frac{1}{i\hbar} {\big[\hat{N}^\alpha,\hat{H}\big]},
\end{equation}
which, for the different components, are
\begin{subequations}\label{eq:singlebondheisenbergoperatorequation}
    \begin{align}
\dv{}{t}\hat{N}^\pm & = \pm iJ\Big[\hat{N}^z\hat{N}^\pm + \hat{N}^\pm \hat{N}^z \\ &  + \hat{N}^z\qty(S(S+1)\hbar^2 + \big(\hat{M}^z\big)^2- \big(\hat{N}^z\big)^2)\Big]\notag \\    
\dv{}{t}\hat{N}^z & = iJ(\hat{N}^+ - \hat{N}^-)\\    
    \dv{}{t}\hat{M}^z & = 0 .   
\end{align}
\end{subequations}
To check consistency, we confirm that, here, the bond magnetization $\hat{M}^z$ commutes with the Hamiltonian. By symmetry, it is thus conserved, as it represents the projection of the total spin onto the $z$-axis.

Up to this point, we have only rewritten the original problem, which contains six (spin) operators with simple commutation relations, in terms of four different operators with more complicated commutation relations. {This transformation by itself does not reduce the complexity of the quantum problem. However, expressing the Hamiltonian in terms of these Néel operators will allow us to introduce a new semiclassical mapping defined at the level of spin correlations. This new mapping derives from the fact that, in constrast to the single-spin algebra, the commutation relations Eqs.~\eqref{eq:single-bond-commutators} of two-spin correlation operators explicitly produce the $\mathfrak{su}(2)$ Casimir $S(S+1)\hbar^2$ which, therefore, also enters in the equations of motion Eqs.~\eqref{eq:singlebondheisenbergoperatorequation}.} Before developing this {mapping}, we extend the formalism by introducing the {correlation operators} on lattices containing multiple bonds.

\subsubsection{Bond-notation for extended lattices}\label{sec:bond-notation}

\begin{figure}[h!]
    \centering
    \includegraphics{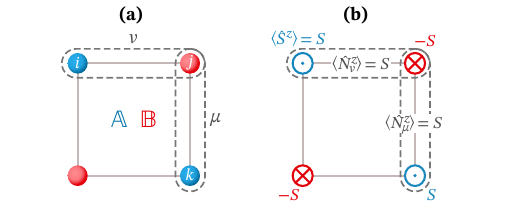}
    \caption{\textbf{(a)} Introduction of the bond-index notation $\nu$ on a bipartite lattice with sublattices $\mathbb{A}$ and $\mathbb{B}$. The bonds are undirected and ordered such that the antiferromagnetic Néel state in \textbf{(b)} has homogeneous $N_\nu^z$ component. In the figure, $\hbar=1$.}
    \label{fig:bond-notation}
\end{figure}

In order to describe the correlation dynamics on an extended bipartite lattice (with sublattices $\mathbb{A}$ and $\mathbb{B}$), containing multiple bonds, we now pass on to using the bond notation $\hat{N}^\alpha \rightarrow \hat{N}^\alpha_\nu$. Each bond $\nu=(i,j)$ is taken as a tuple of site indices $i$ and $j$ [see Fig.~\ref{fig:bond-notation}(a)]. Importantly, different bonds $\nu$ and $\mu$ may share a site/spin. Introducing this bond notation drastically reduces the complexity of the equations in the remainder of this work, since the Néel correlation operators in Eqs.~\eqref{eq:neel-operators} naturally live on the bonds. However, since they are generally not symmetric in the site indices $i$ and $j$, we would also have to consider the inverted bonds $\overline{\nu} = (j, i)$ with $\hat{N}^\alpha_\nu \neq \hat{N}^\alpha_{\overline{\nu}}$. We can do away with these inverted bonds by enforcing a sign structure on the bonds of the lattice. For this, we consider the bonds to be \textit{undirected} $\nu = \{i,j\} = \{j,i\}$ with the \textit{inter-sublattice} correlators consistently defined as
\begin{equation}\label{eq:inter-lattice-correlators}
  \left.\begin{aligned}
  \hat{N}^z_\nu & = \tfrac{1}{2}\big({\hat{S}_i^z-\hat{S}_j^z}\big) \\     
    \hat{M}^z_\nu & = \tfrac{1}{2}\big(\hat{S}_i^z+\hat{S}_j^z\big) \\    
    \hat{N}^\pm_\nu & = \tfrac{1}{2}{\hat{S}_i^\pm \hat{S}_j^\mp}
\end{aligned}\quad  \right\} \quad \text{with}\ i\in\mathbb{A},\ j\in\mathbb{B}
\end{equation}
Under this sign structure, an antiferromagnetic Néel state with sublattices $\mathbb{A}$ ($\mathbb{B}$) polarized in positive (negative) $z$-direction [see Fig.~\ref{fig:bond-notation}(b)] has homogeneous eigenvalues of the $\hat{N}^z_\nu$ operators throughout all inter-sublattice bonds. One can also confirm that the single-bond commutators Eqs.~\eqref{eq:single-bond-commutators} are {the same throughout the lattice in this case}.

\subsubsection{Multi-bond commutation relations} \label{sec:equations-multiplebonds}

In order to derive equations of motion for the Néel operators on multiple bonds ($\nu, \mu, ...$), we also require their mutual commutation relations $[\hat{N}^\alpha_\nu,\hat{N}^\beta_\mu]$. This is where our formulation in terms of bond-wise correlation operators differs drastically from the formulation in terms of individual spins. In particular, while spins at different sites are independent observables, the Néel operators on different bonds can share a spin [Fig.~\ref{fig:bond-notation}(a)] and, therefore, may not commute. We will soon see how these non-zero commutators generate additional correlation operators that enter into the equations of motion. 

Two bonds $\nu$ and $\mu$ can share either zero, one, or two spins. For bonds that do not share any spin, the commutator of the respective correlation operators vanishes trivially. As introduced in Sec.~\ref{sec:bond-notation}, if two bonds share both spins, we identify them as the same undirected bond, so their commutators reduce to the single-bond relations in Eqs.~\eqref{eq:single-bond-commutators}. Finally, if the two bonds share exactly one spin, the commutation relations of the Néel correlation operators can be presented as
\begin{subequations}
    \begin{align}\label{eq:neelmagconnectedbond}
     \big[\hat{N}_\nu^z,\hat{N}_\mu^z\big] &  = 0 \\  \big[\hat{N}_\nu^z,\hat{M}_\mu^z\big] &  = 0 \\  \big[\hat{M}_\nu^z,\hat{M}_\mu^z\big] &  = 0 \\ 
     \big[\hat{N}_\nu^z,\hat{N}_\mu^\pm\big] & = \pm \frac{\hbar}{2}  \hat{N}_\mu^\pm \\ \qty[\hat{M}_\nu^z,\hat{N}_\mu^\pm] & = \pm \frac{\hbar}{2} \varsigma_{\nu\mu} \hat{N}_\mu^\pm\\
     \big[\hat{N}_\nu^\pm,\hat{N}_\mu^\pm\big] & = 0\\
        \big[\hat{N}^+_\nu, \hat{N}^-_\mu\big] & = \frac{\hbar\hat{C}^{\mp}_{\nu\mu}}{2}\Big[\hat{N}^z_\nu + \hat{N}^z_\mu \label{eq:commutator-diagonal-bonds} \\ & \qquad \qquad + \varsigma_{\nu\mu}\big(\hat{M}^z_\nu+\hat{M}^z_\mu\big)\Big]\notag
    \end{align}
\end{subequations}
with 
\begin{equation}\label{eq:def-sigma}
    \varsigma_{\nu\mu}  =\begin{cases}
        1 &  \mathrm{shared}(\nu,\mu) \in \mathbb{A}\\
         -1 &  \mathrm{shared}(\nu,\mu) \in \mathbb{B}
    \end{cases}
\end{equation}
yielding a positive or negative sign depending on whether the shared spin of $\nu$ and $\mu$ is in sublattice $\mathbb{A}$ or sublattice $\mathbb{B}$, and $\hat{C}^+_{\nu\mu}$ and $\hat{C}^-_{\nu\mu}$ being a diagonal-bond operator which we will define soon. {In actuality, the commutator 
\begin{subequations}
    \begin{align}
        \big[\hat{N}^+_\nu, \hat{N}^-_\mu\big] & = \varsigma_{\nu\mu} \hat{C}^\mp_{\nu\mu} \hat{S}^z_j \label{eq:szjgenerated}
    \end{align}
\end{subequations}
generates an $\hat{S}^z_j$ operator at the site $j$ shared by the two bonds $\nu$ and $\mu$. Since we would like to construct a theory directly expressed in terms of the Néel operators, we replace this additional $\hat{S}^z_j$ operator by the Néel vector and magnetization $z$-components on any bonds $\gamma$ that contain $j$ according to
\begin{equation}\label{eq:replace-Sz}
    \hat{S}^z_j = \frac{1}{Z_j}\sum\limits_{\gamma} M^z_\gamma + \varsigma_{\nu\mu} N^z_\gamma
\end{equation}
with $Z_j$ being the number of bonds $\gamma$ \textit{chosen} to replace $\hat{S}^z_j$. To arrive at Eq.~\eqref{eq:commutator-diagonal-bonds}, we have set these replacement bonds to be $\nu$ and $\mu$.
}

\begin{figure}[h!]
    \centering
    \includegraphics{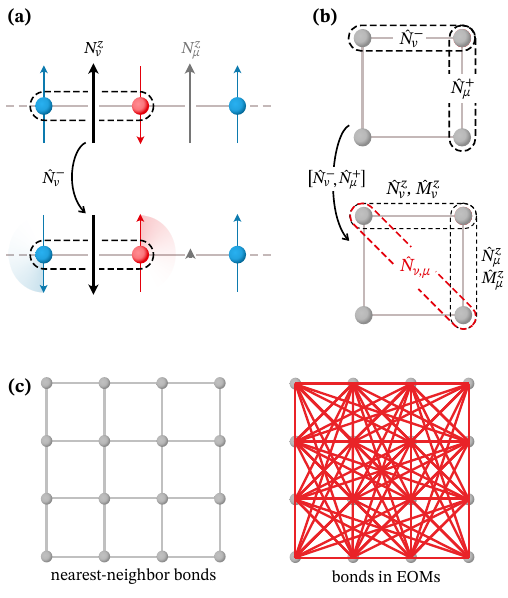}
    \caption{Illustration of how correlation operators propagate across connected bonds due to their non-commutativity. \textbf{(a)} Inverting the Néel vector on a particular bond entails flipping both spins, affecting the Néel vector on the neighboring bond(s). \textbf{(b)} Simultaneous raising and lowering on connected bonds does not commute, generating correlators on diagonal bonds in the process. \textbf{(c)} Therefore, even in a nearest-neighbor model, all possible bonds appear in the equations of motion (EOMs) of the correlators.}
    \label{fig:neel-commutation}
\end{figure}

The commutators Eq.~\eqref{eq:neelmagconnectedbond} reflect the fact that raising/lowering the Néel vector on one bond affects both the Néel vector and magnetization in connected bonds, as is illustrated in Fig.~\ref{fig:neel-commutation}(a). The intuition is clear: Since inverting the Néel vector on a particular bond can only be achieved by flipping both of the corresponding spins, the Néel vector (and magnetization) on all bonds that share these two spins necessarily change.

In addition, simultaneous raising or lowering operations on connected bonds commute with each other. In contrast, raising with $\hat{N}^+_\nu$ on one bond and lowering with $\hat{N}^-_\mu$ on a connected bond do not commute. Instead, as seen in Eq.~\eqref{eq:commutator-diagonal-bonds}, their commutator generates the \textit{intra-sublattice} correlators
\begin{equation}\label{eq:intra-lattice-corr}
\hat{C}^\mp_{\nu\mu} = 
\begin{cases}
   \hat{C}^-_{\nu\mu} =  \tfrac{1}{2}\,\hat{S}_i^+\hat{S}_k^- & \mathrm{shared}(\nu,\mu) \in \mathbb{A},\\[6pt]
   \hat{C}^+_{\nu\mu} = \tfrac{1}{2}\,\hat{S}_i^-\hat{S}_k^+ & \mathrm{shared}(\nu,\mu) \in \mathbb{B},
\end{cases}
\end{equation}
which acts on the diagonal bond connecting the two \emph{outer} spins $i\in\nu$ and $k\in\mu$. These spins are precisely those in the symmetric difference $\nu\,\triangle\,\mu = (\nu \cup \mu) \setminus (\nu \cap \mu)$ of the connected bonds [see Fig.~\ref{fig:neel-commutation}(b)]. The correlators defined by Eq.~\eqref{eq:intra-lattice-corr} live on the bonds connecting sites within the same sublattice and play similar role as the Néel raising/lowering operators $\hat{N}^\pm_\nu$. Their commutation relations with the other operators are analogous and can be found in App.~\ref{App:commsbracks}. 

The non-commutativity of connected correlators has important consequences for the equations of motion. Since commutators between neighboring bonds generate new correlators -- which themselves require equations of motion -- even a Hamiltonian restricted to nearest-neighbor interactions gives rise to correlators on all possible bonds [see Fig.~\ref{fig:neel-commutation}(c)]. As a result, for a connected lattice, the number of coupled equations grows as $\mathcal{O}(n^2)$ with the number of spins $n$ in the system. This shows explicitly that even local interactions necessarily couple to long-range bond correlations, a central feature of our theory.

\subsection{Semiclassical equations of motion}\label{sec:eom-semiclassical}

To this point, we have constructed a set of correlation operators defined on the bonds of a bipartite lattice. In particular, the non-commutativity of connected-bond operators propagates through the whole lattice, generating all-to-all two-point correlation operators in Heisenberg equations of motion
\begin{equation}\label{eq:eom2}
    \dv{{\hat{N}_\nu^\alpha}}{t} = \frac{1}{i\hbar} {\big[\hat{N}_\nu^\alpha,\hat{H}\big]}.
\end{equation}
When kept on the operator level, these equations are exact, but not {closed}. For example, the right-hand side Eq.~\eqref{eq:eom2} may contain the {new} three-point correlation operators $\hat{N}^z_\nu\hat{N}^\pm_\mu$ whose dynamics are to be obtained by taking $[\hat{N}^z_\nu\hat{N}^\pm_\mu,\hat{H}]$. So, generically, the time-evolution of two-point correlations couples to three-point correlations, which couple to four-point correlations and so on. This way, an infinite hierarchy of Heisenberg equations for different static correlation operators is generated. To obtain a tractable set of equations, this hierarchy must be truncated at some level. Truncated equation-of-motion schemes of this type are well established in many-body theories \cite{tyablikovMethodsQuantumTheory1975,jensenRareEarthMagnetism1991,cottamManyBodyTheoryCondensed2020}, for example in the context of $X$-operators \cite{ovcinnikovHubbardOperatorsTheory2004,irkhinSpinOrbitalFerromagnetism2005,mentinkQuantumManybodyDynamics2019}, which are closely related to our spin-correlation operators. In our case, we are able to employ a {specific truncation scheme based on a semiclassical mapping}, which allows us to close the hierarchy at the level of two-point correlations.

In analogy to the classical mapping for single spins $\hat{\boldsymbol{S}}_i$ discussed in Sec.~\ref{sec:roadmap}, we now introduce a mapping from quantum correlation operators $\hat{N}_\nu$ to (semi)classical phase-space functions $\mathbfcal{N}_\nu$ defined on the bonds of the lattice. Recall that, for individual spins, operators are mapped to classical vectors of fixed length $\abs{\mathbfcal{S}_i}=S\hbar$, while the structure of the Heisenberg equations is preserved by replacing commutators with the spin Poisson brackets $\{\cdot,\cdot\}$ defined in Eq.~\eqref{eq:poissonbracket}. 

For spin correlations, the situation is more subtle. As shown in Sec.~\ref{sec:neel-operators}, the commutators of different two-spin correlation operators produce the on-site Casimir $\hat{\boldsymbol{S}}_i^2 = S(S+1)\hbar^2$, which enters explicitly in the corresponding equations of motion. This is not the case for single spins, where replacing $S(S+1) \mapsto S^2$ leaves the structure of the equations unchanged. For correlation operators, however, retaining the exact quantum value of the Casimir--intimately connected to zero-point fluctuations--is essential to reproduce finite-$S$ quantum effects. 

Therefore, we define the semiclassical mapping 
\begin{subequations}\label{eq:semiclassical-limit}
    \begin{align}
    \hat{\boldsymbol{N}}_\nu,\, \hat{H} \ \rightsquigarrow\   & \mathbfcal{N}_\nu,\, \mathcal{H}_\mathrm{sc},\\
    \frac{1}{i\hbar}\big[\hat{\boldsymbol{N}}_\nu,\hat{H}\big] \ \rightsquigarrow\ & \big\{\mathbfcal{N}_\nu,\mathcal{H}_\mathrm{sc}\big\},\\
    & \text{with}\ \abs{\mathbfcal{S}_i}^2 = S(S+1)\hbar^2. \label{eq:replace_spin_value}
    \end{align}
\end{subequations}
This mapping associates quantum correlation operators with the semiclassical correlation components
\begin{equation}
    \mathbfcal{N}_\nu = \mqty[\mathcal{N}^+_\nu,& \mathcal{N}^-_\nu,& \mathcal{N}^z_\nu,& \mathcal{M}^z_\nu],
\end{equation}
while preserving both the single-spin Casimir and the algebraic structure of the equations of motion. Here and below, we use the arrow ``$\rightsquigarrow$'' to distinguish our new semiclassical mapping from the conventional single-spin semiclassical theory. 

The Poisson brackets $\{\mathcal{N}^\alpha_\nu,\mathcal{N}^\beta_\mu\}$ between different correlation components follow directly from the original commutators through the mapping \eqref{eq:semiclassical-limit}. Equivalently, they can be obtained by expressing the semiclassical correlations in terms of classical spin components, applying the spin Poisson brackets according to Eq.~\eqref{eq:poissonbracket}, and replacing $\mathcal{S}^2$ with $S(S+1)\hbar^2$. After this mapping, the equations of motion of the semiclassical correlations take the Hamiltonian form
\begin{equation}
    \dv{\mathbfcal{N}_\nu}{t} = \qty{\mathbfcal{N}_\nu, \mathcal{H}_\mathrm{sc}}.
\end{equation}
The semiclassical mapping can also be viewed as a decoupling of expectation values, but now at the level of{two-point correlations} rather than single-spin observables. Taking the expectation value of Eq.~\eqref{eq:eom2} leads to closed equations for the components of $\langle{\hat{\boldsymbol{N}}_\nu}\rangle \approx \mathbfcal{N}_\nu$ only if higher-order terms are factorized, 
\begin{align}\label{eq:decoupling_new}
    \big\langle \hat{N}^z\hat{N}^\pm \big\rangle & \rightsquigarrow \big\langle{\hat{N}^z} \big\rangle\big\langle {\hat{N}^\pm}\big\rangle,\\
    \big\langle\hat{N}^z(\hat{M}^z)^2 & \rightsquigarrow \big\langle\hat{N}^z \big\rangle\big\langle \hat{M}^z\big\rangle^2,
\end{align}
and so on. 

\begin{figure}[h!]
    \centering
    \includegraphics{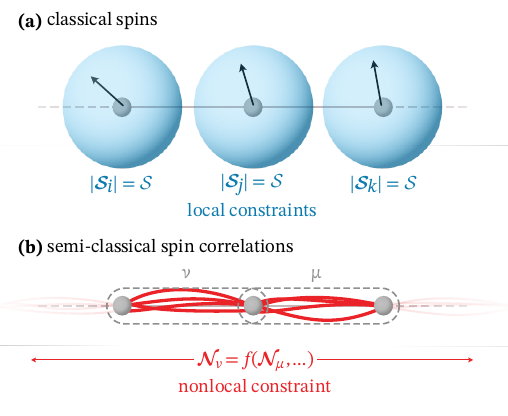}
    \caption{Illustration of the different constraints on classical spins $\mathbfcal{S}_i$ and semiclassical correlations $\mathbfcal{N}_\nu=[\mathcal{N}^+_\nu,\mathcal{N}^-_\nu,\mathcal{N}^z_\nu,\mathcal{M}^z_\nu]$. \textbf{(a)} While classical spins are constrained independently of each other due to the commutation of different spin operators, \textbf{(b)} correlation operators on different bonds $\nu$ and $\mu$ generally do not commute and therefore constrain each other in the semiclassical limit.}
    \label{fig:nonlocal_constraint}
\end{figure}

It is important to note that semiclassical correlations are inherently \emph{non-local} objects. Unlike classical spins, which satisfy a \textit{local} constraint $\abs{\mathbfcal{S}_i}=\mathcal{S}$ as a direct consequence of the local rotational symmetry, the values of the correlation components on a given bond are constrained by all (directly or indirectly) connected bonds via the underlying commutation relations. This is illustrated schematically in Fig.~\ref{fig:nonlocal_constraint}. Intuitively, the phase space of $n$ classical spins can be viewed as the Cartesian product of $n$ spheres. Rewriting this phase space in terms of correlation variables distributes the $\mathfrak{su}(2)$ constraint across the entire lattice. This non-local structure is a fundamental feature of the semiclassical mapping at the level of correlations.

In summary, by {mapping a quantum to a classical theory} on the level of the Néel correlation operators, we obtained closed semiclassical equations of motion for spin correlations that can be expressed in terms of Poisson brackets. In the following, we will {augment this semiclassical mapping} by introducing dissipation, obtaining a correlation-level analogue of the Landau–Lifshitz–Gilbert equation.

\subsection{Dissipative state preparation and damping}\label{sec:damping}

Apart from being able to describe the dynamics of a spin system, it is essential, in many situations, to also be able to determine equilibrium states or even the ground state of a system. In this section, we achieve this by deriving dissipative terms in the semiclassical equations of motion for the spin correlations. Using the dissipative bracket \cite{blochEulerPoincareEquationsDouble1996,liuMATHEMATICALREVISIONLANDAULIFSHITZ2009}, a classical analogue of the quantum-mechanical anticommutator, we will be able to construct a correlation-level analog of the Landau-Lifshitz-Gilbert equation for spins.

In classical magnetism, equilibrium states represent stable fixed points in phase space at which all classical torques vanish, $\partial_t \mathbfcal{S}_i=0$, and the classical Hamilton function $\mathcal{H}_\mathrm{cl}(\mathbfcal{{S}}_1, ..., \mathbfcal{{S}}_n)$ exhibits a local minimum. Such states are commonly obtained by implementing dissipative terms such as Gilbert damping into the equations of motion, which, subject to the constraint $\abs{\mathbfcal{S}_i}=\mathcal{S}$, let the system evolve to a state with vanishing torque. As discussed in Sec.~\ref{sec:eom-semiclassical}, while this constraint is local for single-spin observables, it becomes \textit{nonlocal} for spin correlations. In both cases, it is directly encoded into the respective Poisson brackets that generate the corresponding Hamiltonian dynamics. 

A first intuition of how to introduce dissipation in the Hamiltonian formalism comes from quantum mechanics: While, in the Heisenberg picture, conservative dynamics are generated by the usual commutator $[\cdot,\cdot]$, the dissipative evolution of a quantum system is connected to the \textit{anticommutator} $[\hat{A},\hat{B}]_+ = \hat{A}\hat{B} + \hat{B}\hat{A}$. This suggests that implementing dissipative terms for semiclassical spin correlations in a Hamiltonian picture requires a different algebraic structure from the Poisson bracket $\{\cdot,\cdot\}$. Indeed, in classical systems, this role is played by the \textit{dissipative bracket} $\{\cdot,\cdot\}_+$ that, in terms of classical spin vectors, can be given as
\begin{equation}\label{eq:dissipative-bracket}
        \qty{\mathcal{F},\mathcal{G}}_+=  \sum\limits_{j=1}^n  \qty({\mathbfcal{S}}_j\times \pdv{\mathcal{F}}{{\mathbfcal{S}}_j}) \cdot \qty({\mathbfcal{S}}_j\times \pdv{\mathcal{G}}{{\mathbfcal{S}}_j})
\end{equation} 
for two functions $\mathcal{F}$ and $\mathcal{G}$ on the classical phase space. One can show that this form is bilinear and satisfies the usual product rule, but, in contrast to the Poisson bracket, it is symmetric in its arguments. In this sense, it is a classical version of the quantum-mechanical anticommutator $\qty[\cdot,\cdot]_+$ and generates dissipative dynamics in the classical phase space. Using this bracket, the classical dissipative equation of motion at the spin level can be written as
\begin{subequations}\label{eq:llg-bracket}
    \begin{align}
%\begin{split}
        \dv{}{t}\boldsymbol{S}_i & = \qty{\boldsymbol{S}_i,\mathcal{H}_\mathrm{cl}} - \eta \qty{\boldsymbol{S}_i,\mathcal{H}_\mathrm{cl}}_+\\
    & = \mathbfcal{S}_i \times \pdv{\mathcal{H}_\mathrm
    {cl}}{ \mathbfcal{S}_i} + \eta\mathbfcal{S}_i \times \qty[ \mathbfcal{S}_i \times \pdv{\mathcal{H}_\mathrm
    {cl}}{ \mathbfcal{S}_i}],
%\end{split}
\end{align}
\end{subequations}
which is consistent with the Landau-Lifshitz equation with a phenomelogical damping term.
Equipped with the dissipative bracket, describing damping in terms of semiclassical spin correlations is straightforward. All we need to do is transcribing Eq.~\eqref{eq:llg-bracket} to semiclassical correlations $\mathbfcal{N}_\nu$ by computing their dissipative brackets and replace any occurences of $S$ with $S(S+1)$ in accordance with the semiclassical mapping Eq.~\eqref{eq:semiclassical-limit}. These brackets follow immediately from the dissipative brackets between underlying spins at sites $i$ and $j$, which are obtained from Eq.~\eqref{eq:dissipative-bracket} as
\begin{align}\label{eq:spin-dissipative-bracket}
    \big\{\mathcal{S}^\alpha_i,\mathcal{S}^\beta_j\big\}_+ & = \delta_{ij}\begin{cases} -\mathcal{S}^\alpha_i\mathcal{S}^\beta_j & \alpha \neq \beta\\
    -(\mathcal{S}_i^\alpha)^2 + S(S+1)\hbar^2 & \alpha = \beta \end{cases}.
\end{align}
Here, we have already replaced $\mathcal{S}=S(S+1)\hbar^2$ inline with the semiclassical mapping. Notice how, in contrast to the antisymmetric Poisson bracket (or the commutator), the dissipative bracket of a spin component with itself is non-zero, \textit{e.g.}  $\qty{{S}^x,{S}^x}_+\neq 0$. Using Eq.~\eqref{eq:spin-dissipative-bracket} and the properties of the dissipative bracket, one can obtain $\{ \mathcal{N}_\nu^\alpha, \mathcal{N}^\beta_\mu\}_+$ for all correlation components on an extended lattice. For example, on a single bond, one finds
\begin{subequations}
    \begin{align}
    \qty{\mathcal{N}_\nu^z,\mathcal{M}_\nu^z}_+ & = - \mathcal{N}_\nu^z\mathcal{M}_\nu^z\\
        \qty{\mathcal{N}_\nu^z,\mathcal{N}_\nu^\pm}_+ & = - \mathcal{N}_\nu^\pm \mathcal{N}_\nu^z\\
\qty{\mathcal{N}_\nu^z,\mathcal{N}_\nu^z}_+ & = - \frac{ \qty(\mathcal{N}_\nu^z)^2 +  \qty(\mathcal{M}_\nu^z)^2}{2} + \frac{S(S+1)\hbar^2}{2}\\
          &\ \,  \vdots \notag 
\end{align} %    \qty{\mathcal{M}_\nu^z,\mathcal{M}_\nu^z}_+ = 
\end{subequations}
The complete list of brackets is presented in the App.~\ref{App:commsbracks}. With that, we can present the Landau-Lifshitz-Gilbert analogue of semiclassical spin-correlation dynamics as 
\begin{equation}\label{eq:spincorrllg}
       \dv{\mathbfcal{N}_\nu}{t} = \qty{\mathbfcal{N}_\nu,\mathcal{H}_\mathrm{sc}} -\eta \qty{\mathbfcal{N}_\nu, \mathcal{H}_\mathrm{sc}}_+ 
\end{equation}
%
%To this point, we have not derived a general form akin to the spin equation involving an effective magnetic field $\partial \mathcal{H}/\partial \mathbfcal{S}$. 
Their explicit form depends on the particular Hamiltonian. For example, in the case of an isolated dimer with Heisenberg exchange discussed before, one has (with bond index $\nu$ omitted)
\begin{widetext}
    \begin{subequations}
    \begin{align}
        \dv{}{t}\mathcal{N}^\pm & = \qty{\mathcal{N}^\pm,\mathcal{H}_\mathrm{sc}}+ 2\eta  \mathcal{N}^\pm \mathcal{H}_\mathrm{sc} - \eta J S(S+1)\hbar^2\qty[S(S+1)\hbar^2 -\qty(\mathcal{M}^z)^2 -\qty(\mathcal{N}^z)^2 ]
 \\    
\dv{}{t}\mathcal{N}^z & = \qty{\mathcal{N}^z,\mathcal{H}_\mathrm{sc}} + \eta \mathcal{N}^z \left[\mathcal{H}_\mathrm{sc} + JS(S+1) \hbar^{2}\right]\\
\dv{}{t}\mathcal{M}^z & = \underbrace{\qty{\mathcal{M}^z,\mathcal{H}_\mathrm{sc}}}_{0} + \eta \mathcal{M}^z \left[\mathcal{H}_\mathrm{sc} - JS(S+1) \hbar^{2}\right]\label{eq:mag-dissipation}
    \end{align}
\end{subequations}
\end{widetext}
where the Hamiltonian dynamics are kept in terms of the Poisson brackets and the semiclassical energy is
\begin{equation}
    \mathcal{H}_\mathrm{sc} = J\qty[\mathcal{N}^+ + \mathcal{N}^- + \qty(\mathcal{M}^z)^2 - \qty(\mathcal{N}^z)^2],
\end{equation}
which is \textit{not conserved} here. Equation~\eqref{eq:mag-dissipation} shows that if dissipation is present, the magnetization of the isolated dimer is generally not conserved. The additional dissipative terms naturally drive the system to a local equilibrium in terms of correlations.

%\st{Even though the mutual commutation relations of Néel correlation operators on connected bonds induce a nonlocal constraint at the semiclassical level that makes straightforward energy minimization or Legendre transformation to a Lagrangian formalism cumbersome, we have seen that this obstruction can be circumvented using the dissipative bracket.}

In summary, our semiclassical mapping provides a natural way to accomodate for phenomenological damping based on the dissipative bracket.
Using this algebraic structure, we have now derived equations of motion for semiclassical spin correlations, which can describe both conservative and dissipative dynamics. %\st{This, in particular, allows us to obtain equilibrium states in terms of semiclassical spin correlations.} 

\subsection{Calculation of full correlation function}\label{sec:full-correlation}

Lastly, to assess magnetic ordering and dynamics thereof, it is often desired to calculate the full {bond-wise} equal-time correlation function $\mathcal{C}_\nu = \langle \hat{\boldsymbol{S}}_i\cdot\hat{\boldsymbol{S}}_j \rangle$  between \textit{any} two lattice sites $i$ and $j$ sharing a bond $\nu$. For this, the expectation values of $\langle \hat{{S}}^x_i\hat{{S}}^x_j + \hat{{S}}^y_i\hat{{S}}^y_j\rangle = 2\,\mathrm{Re}\,\mathcal{N}^{(\pm)}_{\nu}$ are directly obtained by the semiclassical multi-bond equations of motion which explicitly yield the expectation values $\mathcal{N}^\pm_\nu$ and $\mathcal{N}_\nu$ of inter and intra-lattice correlators defined according to Eq.~\eqref{eq:inter-lattice-correlators} and Eq.~\eqref{eq:intra-lattice-corr}. However, the correlation $\langle \hat{{S}}^z_i\hat{{S}}^z_j\rangle$, where $i$ and $j$ are not nearest neighbors, is not directly determined as the Néel vector and magnetization components are only defined on the nearest-neighbor bonds. It can instead be obtained by replacing the $\hat{S}^z$ operators according to Eq.~\eqref{eq:replace-Sz} with the Néel vector and magnetization $z$-components on the adjacent bonds and subsequently decoupling the expectation values according to Eq.~\eqref{eq:decoupling_new} {inline with the semiclassical mapping}. As a result, one obtains
\begin{widetext}
    \begin{equation}\label{eq:full-correlation-sc}
    \mathcal{C}_\nu = 2 \,\mathrm{Re}\,\mathcal{N}^{(\pm)}_\nu + \frac{1}{Z_iZ_j}\sum\limits_{\nu,\mu} \mathcal{M}^z_\nu \mathcal{M}^z_\mu +  \varsigma_i\mathcal{N}^z_\nu \mathcal{M}^z_\mu + \varsigma_j\mathcal{M}^z_\nu \mathcal{N}^z_\mu + \varsigma_i\varsigma_j\mathcal{N}^z_\nu \mathcal{N}^z_\mu
\end{equation}
\end{widetext}
where, here, $\nu$ ($\mu$) runs over all $Z_i$ ($Z_j$) nearest-neighbor bonds that contain the site $i$ ($j$) and $\varsigma_{i,j} = \pm 1$ depending on whether the respective site is in sublattice $\mathbb{A}$ or $\mathbb{B}$.

\section{Application to Heisenberg antiferromagnets}

In the previous Sections, we have established the theory of semiclassical spin correlations, including their Hamiltonian and dissipative dynamics. We now proceed to compare its predictions to entirely classical and fully quantum magnetism in simple situations for which exact solutions are feasible. In particular, we will investigate the ground-state properties and quantum oscillations present in Heisenberg antiferromagnets. Despite their relatively simple structure, these systems feature an abundance of genuinely quantum properties such as highly entangled ground states, two-magnon modes \cite{elliottEffectsMagnonmagnonInteraction1969,kamraAntiferromagneticMagnonsHighly2019,bossiniLaserdrivenQuantumMagnonics2019,formisanoCoherentTHzSpin2024a,bassantEntangledMagnonpairGeneration2024, azimi-mousolouMagnonShakeupEntanglement2025}, or supermagnonic propagation \cite{fabianiSupermagnonicPropagationTwoDimensional2021,boumanTimedependentSchwingerBoson}. The following examples will demonstrate how semiclassical spin correlations effectively interpolate between classical and quantum magnetism, while capturing dynamics that are simultaneously nonlinear and nonclassical. To this end, we consider the Heisenberg Hamiltonian and its semiclassical counterpart, written in terms of Néel-correlation operators
\begin{subequations}
    \begin{align}
            \hat{H} & = J\sum\limits_\nu \qty[\hat{N}^+_\nu + \hat{N}^-_\nu + \big(\hat{M}^z_\nu\big)^2 - \big(\hat{N}^z_\nu\big)^2]\\
    \rightsquigarrow \mathcal{H}_\mathrm{sc} &  = J\sum\limits_\nu \qty[\mathcal{N}^+_\nu + \mathcal{N}^-_\nu + \big(\mathcal{M}^z_\nu\big)^2 - \big(\mathcal{N}^z_\nu\big)^2],
    \end{align}
\end{subequations}
with $J>0$. Here $\nu$ runs over all nearest-neighbor bonds without double-counting (as we have fixed a sign structure).

\subsection{Semiclassical groundstates}\label{sec:ground-states}

Before investigating any dynamics, we will have a look at the properties of the ground states in the semiclassical limit. In classical magnetism, the total energy in the antiferromagnetic Heisenberg model is famously minimized by a degenerate set of Néel states for which all spins on the two sublattices $\mathbb{A}$ and $\mathbb{B}$ are antiparallel, 
\begin{equation}
    E_\mathrm{cl}\equiv \mathcal{H}_\mathrm{cl}(\mathbfcal{S}, -\mathbfcal{S},\mathbfcal{S},-\mathbfcal{S},...) = -JS^2\hbar^2n_\mathrm{b} 
\end{equation}
with $n_\mathrm{b}$ being
the number of bonds. In quantum theory, the Néel state corresponds to a simple product state with only trivial correlations. By contrast, the true ground state of the Heisenberg antiferromagnet is entangled and therefore cannot be described within a purely classical framework. The quantum mechanical groundstate energy is correspondingly lower than the classical value. Since our approach is formulated in terms of two-point correlations, it is natural to ask to what extent it can account for these quantum deviations from classical theory. %\st{Here, for better comparison,  we start with small systems, where the quantum Hamiltonian can still be diagonalized exactly.} 
Below, we compare the results of our semiclassical theory to exact diagonalization. The semiclassical minima in terms of the spin correlations are obtained by starting in the Néel state and relaxing the system with the dissipative equation of motion Eq.~\eqref{eq:spincorrllg} with damping parameter $\eta = 0.1$.

\subsubsection{Single bond}

\begin{figure}[h!]
    \centering
    \includegraphics{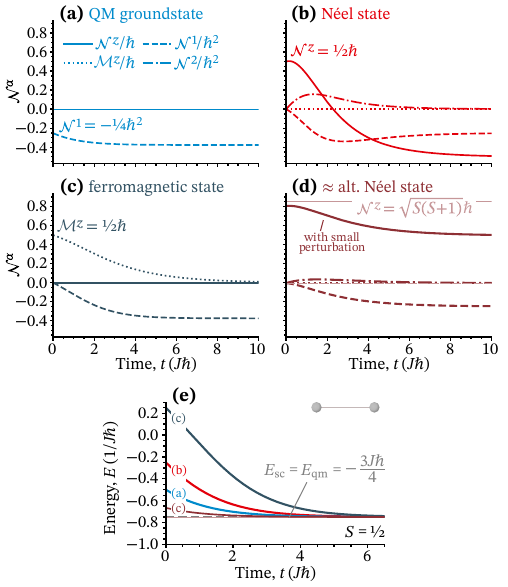}
    \caption{Relaxation dynamics of different initial states in the semiclassical limit, all evolving to degenerate semiclassical ground states. \textbf{(a-d)} show the Néel-correlation components for different initial states evolved according to the dissipative equations of motion Eq.~\eqref{eq:spincorrllg} of a single antiferromagnetically coupled dimer with $S=\sfrac{1}{2}$ and damping parameter $\eta = 0.1$. The corresponding energies during the relaxation are shown in \textbf{(e)}.}
    \label{fig:gs-relaxation}
\end{figure}

In order to understand the nature of ground states in terms of semiclassical spin correlations, we will first focus on the case of an isolated antiferromagnetic dimer, one of the simplest cases where nonclassical effects dominate. The ground state is an entangled singlet, which is a superposition of classical product states. For the lowest spin values $S$, the exact ground states are given by the following singlet states (written in the $\ket{m_1,m_2}$ basis)
\begin{widetext}
    \begin{align}\label{eq:singlet-state-qm}
    S=\sfrac{1}{2}\qquad E_\mathrm{qm} & = - \frac{3J\hbar^2}{4} &  \ket{\psi_\mathrm{g}} & = \frac{\ket{\tfrac{1}{2},-\tfrac{1}{2}} - \ket{-\tfrac{1}{2},\tfrac{1}{2}}}{\sqrt{2}}\\
    S=1\qquad E_\mathrm{qm} & = - {2J\hbar^2} &  \ket{\psi_\mathrm{g}} & = \frac{\ket{1,-1} - \ket{0,0} + \ket{-1,1}}{\sqrt{3}}\\
     S =\sfrac{3}{2}\qquad E_\mathrm{qm} & = - \frac{15J\hbar^2}{4} & \ket{\psi_\mathrm{g}} &= \frac{\ket{\tfrac{3}{2},-\tfrac{3}{2}} +\ket{\tfrac{1}{2},-\tfrac{1}{2}} - \ket{-\tfrac{1}{2},\tfrac{1}{2}}- \ket{-\tfrac{3}{2},\tfrac{3}{2}}}{2}
\end{align}
\end{widetext}
with the energies for any $S$ being
\begin{equation}
    E_\mathrm{qm} = - S(S+1)J\hbar^2 < E_\mathrm{cl}
\end{equation} 
for the case of a single bond.
The expectation values $\expval{...}_\mathrm{g}$ of the Néel correlation components in the quantum-mechanical ground states $\ket{\psi_\mathrm{g}}$ are (with bond index omitted)
\begin{equation}
\mathbfcal{N}_\mathrm{qm} \equiv \qty[\big\langle \hat{N}^+ \big\rangle_\mathrm{g}, \big\langle{\hat{N}^-}\big\rangle_\mathrm{g}, \big\langle{\hat{N}^z}\big\rangle_\mathrm{g}, \big\langle{\hat{M}^z}\big\rangle_\mathrm{g}]
\end{equation}
with 
\begin{equation}
\big\langle {\hat{N}^\pm} \big\rangle_\mathrm{g}= -\frac{1}{3}S(S+1)\hbar^2, \ \big\langle {\hat{N}^z}\big\rangle_\mathrm{g}=\big\langle {\hat{M}^z}\big\rangle_\mathrm{g}=0
\end{equation}
where the non-zero expectation values  $\langle\hat{N}^\pm\rangle_\mathrm{g}$ are a reflection of the fact that $\ket{\psi_\mathrm{g}}$ is not an eigenstate of $\hat{N}^z$. By inserting these correlation expectation values $\mathbfcal{N}_\mathrm{qm}$ into the semiclassical Hamiltonian, one can quickly check that this state has a higher semiclassical energy  
\begin{equation}
\quad\mathcal{H}_\mathrm{sc}\qty(\mathbfcal{N}_\mathrm{qm}) = -\frac{2}{3}S(S+1)J\hbar^2\ >\ E_\mathrm{qm}.
\end{equation}
This indicates that the semiclassical mapping, which replaces $\langle(\hat{N}^z)^2\rangle$ by $\langle\hat{N}^z\rangle^2$, yields an energy that is higher than the quantum-mechanical groundstate energy. However, interestingly, the correlation expectation values $\mathbfcal{N}_\mathrm{qm}$ of this state are \textit{not} an equilibrium solution of the 
semiclassical Hamiltonian equations,
\begin{equation}
    \dv{\mathbfcal{N}_\mathrm{qm}}{t} = \qty{\mathbfcal{N}_\mathrm{qm}, \mathcal{H}_\mathrm{sc}} \neq 0.
\end{equation}
Indeed, Fig.~\ref{fig:gs-relaxation}(a) and Fig.~\ref{fig:gs-relaxation}(e) the Néel variables relax to lower energy $E_\mathrm{sc}$ when evolved under the dissipative equation of motion Eq.~\eqref{eq:spincorrllg} with $\eta = 0.1$. Here, instead of the complex components $\mathcal{N}^\pm$, we plot their real counterparts $\mathcal{N}^{(1)} \equiv (\mathcal{N}^+ +\mathcal{N}^-)/2$ and $\mathcal{N}^{(2)} \equiv (\mathcal{N}^+ -\mathcal{N}^-)/(2i)$.

\begin{figure}[h!]
    \centering
    \includegraphics{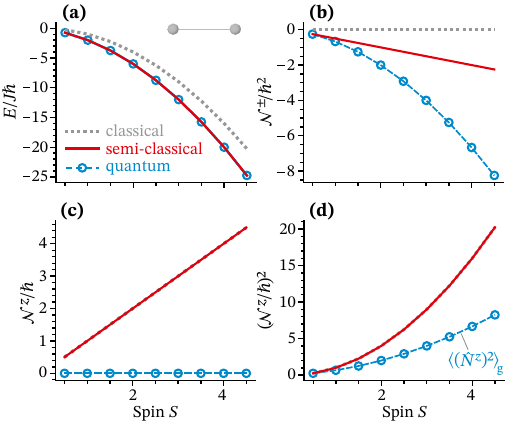}
    \caption{\textbf{(a)} Energy and \textbf{(b-d)} Néel expectation values of the ground states of a single antiferromagnetically coupled bond in the quantum, semiclassical, and fully classical case for increasing spin quantum number $S$. In all cases, the magnetization $\mathcal{M}^z=0$.}
    \label{fig:gs-single-bond}
\end{figure}

This new semiclassical ground state exhibits the correct energy of the singlet state 
\begin{equation}
    E_\mathrm{sc} = E_\mathrm{qm}
\end{equation}
for any $S$. This is shown in Fig.~\ref{fig:gs-single-bond}(a), which presents a comparison of the quantum, semiclassical, and entirely classical energy dependence on $S$. However, in contrast to the quantum ground state, but much like the classical Néel state, the semiclassical ground state is degenerate. To illustrate this, Fig.~\ref{fig:gs-relaxation} shows the relaxation from different initial states into several degenerate ground states with non-zero Néel vector, which are connected to each other by rotations in spin space. A representative equilibrium of the semiclassical equations is given by
\begin{align}
    \mathcal{N}^\pm & = -S/2\hbar^2  & \mathcal{N}^z& =S\hbar &  \mathcal{M}^z&=0.
\end{align}
Notice that this semiclassical groundstate also exhibits non-zero $\mathcal{N}^\pm$ components much like in the quantum theory (though with different $S$ scaling). However, while the singlet state features $\langle \hat{N}^z\rangle  = 0$, the semiclassical groundstate exhibits a non-zero Néel component, $\mathcal{N}^z=S\hbar$, thereby breaking this symmetry in a similar way as the classical equilibrium. These results are summarized in Fig.~\ref{fig:gs-single-bond}. It is worth noting that although the quantum groundstate has a vanishing expectation value of $\hat{N}^z$, it still exhibits finite fluctuations, reflected in a nonzero expectation value of $(\hat{N}^z)^2$ which, similar to the (semi)classical cases, grows with $S$, though with a smaller rate [see Fig.~\ref{fig:gs-single-bond}(d)]. These results show that the ground state of a single bond in the semiclassical limit has the correct energy, but its spin correlations exhibit values in between the quantum and the entirely classical case. 

Once our semiclassical mapping associates $\mathcal{S}^2 = S(S+1)\hbar^2$, it is worth noting that the ground state in terms of the semiclassical spin correlations is \textit{not} simply the ground state of two classical spins with increased length. This would correspond to a Néel state with $\mathcal{M}^z=\mathcal{N}^\pm = 0$ and $\mathcal{N}^z=\sqrt{S(S+1)}\hbar$. Although this state has the correct semiclassical energy $\mathcal{H}_\mathrm{sc}=-S(S+1)J\hbar^2$ and is indeed a fixed point of the dynamic equations, it is unstable with respect to small perturbations. It is also worth noting that initial values of $\mathcal{N}^z > \sqrt{S(S+1)}\hbar$ lead to a rapid blow-up of the solution. This is consistent with the fact that such semiclassical configurations corresponding to spin projections outside of the sphere with radius 
$\sqrt{S(S+1)}\hbar$ are unphysical.

\subsubsection{Multiple bonds}

\begin{figure}[h!]
    \centering
    \includegraphics{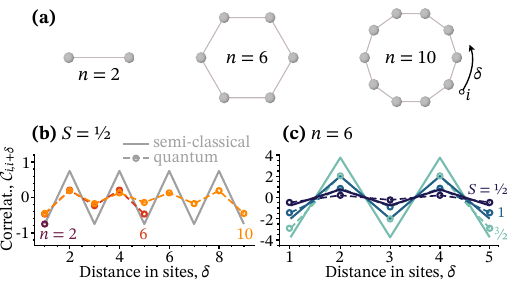}
    \caption{Comparison of the semiclassical groundstate correlations with exact diagonalization of the quantum Hamiltonian. The curves show \textbf{(a)} one-dimensional antiferromagnetic Heisenberg spin chains with periodic boundary conditions \textbf{(b)} for different numbers of spins $n$ and \textbf{(c)} different spin quantum numbers $S$ of the spins in the chains.}
    \label{fig:correlation-1d-chain}
\end{figure}

By investigating the semiclassical ground state of a single bond, we found that it features qualities between the fully quantum and the entirely classical case. It is instructive to ask how this situation changes in the case of more than one bond. As we depart from this simple case, all-to-all correlation components will be become relevant (see Sec.~\ref{sec:equations-multiplebonds}). By relaxing from the Néel state in several one and two-dimensional lattices with $n_\mathrm{b}$ nearest-neighbor bonds, with and without periodic boundary conditions, we find that 
\begin{equation}
    E_\mathrm{sc} = -S(S+1)h^2J n_\mathrm{b} < E_\mathrm{cl}
\end{equation}
which is a multiple of the single-bond energy (that coincided with the energy of the quantum-mechanical groundstate). This multiplicative property is similar to the classical Heisenberg antiferromagnet, and is indeed an underestimation of the exact quantum-mechanical value. In the actual quantum system containing one or more bonds, the true ground-state energy $E_\mathrm{qm}$ satisfies \cite{liebClassicalLimitQuantum1973}
\begin{equation}
    E_\mathrm{sc} \leq E_\mathrm{qm} < E_\mathrm{cl}
\end{equation}
where the equality only holds for $n_\mathrm{b}=1$ bonds (as we have seen in the previous section). For more than one bond, the quantum system has a higher energy than the semiclassical value $E_\mathrm{sc}$, as equality would require each spin to be in a singlet state (maximally entangled) with each of its nearest neighbors. This, however, fails to be an eigenstate for the extended lattice.. Because of this, the scaling of $E_\mathrm{qm}$ as a function of system size and dimension is distinct from that of $E_\mathrm{sc}$. To illustrate this, we calculate the static correlation functions both in the semiclassical limit and from exact diagonalization in the respective ground states of one-dimensional spin chains with periodic boundary conditions, schematically shown in Fig.~\ref{fig:correlation-1d-chain}(a). The comparison with exact diagonalization in Fig.~\ref{fig:correlation-1d-chain}(b) for $S=\sfrac{1}{2}$ shows that, although the semiclassical theory captures the oscillatory behavior of the correlation function in its antiferromagnetic ground state, it does not recover the decrease in correlation with distance or the dependence on the system size. Indeed, for the full semiclassical correlation  according to Eq.~\eqref{eq:full-correlation-sc} one obtains the singlet value on all bonds
\begin{equation}
    \mathcal{C}_\nu = \pm S\qty(S+1)\hbar^2
\end{equation}
with a positive or negative sign depending on whether the bond $\nu$ connects spins in the same sublattice or not. Figure~\ref{fig:correlation-1d-chain}(c) shows a comparison of the scaling behavior of the correlation function with $S$ for a fixed number of spins. Due to the different single-bond scaling between the exact quantum ground state and the semiclassical limit, the latter may over- or underestimate the absolute value of the full correlation.

In summary, the semiclassical theory reproduces the correct ground-state energy for a single bond but does not completely capture correlations beyond that. In fact, for extended systems containing connected bonds, the semiclassical energy extends below the exact quantum groundstate energy.

\subsection{Nonlinear quantum oscillations in the Néel state}\label{sec:neel-oscillations}

After shedding light on the ground-state properties of Heisenberg antiferromagnets based on semiclassical spin correlations, we now investigate their dynamical behavior. Since we derived the equations of motion for spin correlations without any linearization at the level of the original operators, it is natural to ask if our theory can capture the nonlinear dynamics of their expectation values. Indeed, we will see that our semiclassical theory can describe dynamics that are both nonlinear and nonclassical. 

An elementary example that features nontrivial dynamics of quantum correlations is the oscillations in the antiferromagnetic Néel state. While the Néel state represents an equilibrium state of the classical Heisenberg antiferromagnet (where all $\partial_t\mathbfcal{S}_i=0$), it is not even an eigenstate of the quantum Hamiltonian and will fluctuate, indicated by oscillations in the expectation value of the Néel operators. These oscillations are related to the magnon zero-point fluctuations typically encountered in the ground state of antiferromagnets. Despite being a very simplified example, capturing the evolution of states that are not energy eigenstates is of particular relevance for the study of quantum systems, for example, within ultrafast magnetism. In this context, a spin system can be prepared in the Néel state $\ket{\text{Néel}}$ by a quench of the anisotropic exchange interaction from an antiferromagnetic Ising Hamiltonian to the Heisenberg Hamiltonian,
\begin{equation}
 \begin{split}
         \hat{H} & = \underbrace{J\sum\limits_\nu \qty[\big(\hat{M}^z_\nu\big)^2 - \big(\hat{N}^z_\nu\big)^2]}_{\hat{H}_\mathrm{Ising}}\ \\  & \qquad\qquad + \Theta(t)\ J\sum\limits_\nu \qty[\hat{N}^+_\nu + \hat{N}^-_\nu],
 \end{split}
\end{equation}
with $\Theta(t)$ being the step function that is zero for $t<0$ and one for $t>0$. Initially, the system is described only by the Ising Hamiltonian for which the Néel state is a groundstate also for quantum spins. Once the system is quenched at $t=0$ to feature the full Heisenberg Hamiltonian, the Néel state evolves as a superposition of energy eigenstates of the new Hamiltonian $\ket{\phi_i}$ according to
\begin{equation}
    \ket{\psi(t)} = \sum\limits_i c_i e^{-i\Omega_it} \ket{\phi_i} %+ c_2 e^{-i\Omega_2t} \ket{\phi_2} +  ... % c_3 e^{-i\Omega_3t} \ket{\phi_3} + 
\end{equation}
with complex coefficients $c_i=\braket{\text{Néel}}{\phi_i}$ and eigenfrequencies $\Omega_i$. Due to the interference of different energy eigenstates, the expectation values of the Néel-operator components 
\begin{equation}\label{eq:general-qm-neel-oscillation-expval}
    \big\langle{\hat{\boldsymbol{N}}_\nu(t)}\big\rangle = \sum_{\omega_m}  {\boldsymbol{N}}_\nu(\omega_m) e^{- i\omega_m t}
\end{equation}
will feature an abundance of frequency components $\omega_m$, with Fourier coefficients ${\boldsymbol{N}}_\nu(\omega_m)$. These frequencies are determined by combinations of the eigenfrequencies of the Hamiltonian, \textit{e.g.} $\Omega_1 - \Omega_2$, $\Omega_2 + \Omega_3$ and so on. Here, we write $\hat{\boldsymbol{N}} = [\hat{N}^+,\hat{N}^-,\hat{N}^z,\hat{M}^z]$ for brevity, to distinguish it from the semiclassical correlations $\mathbfcal{N}$. Depending on the eigenspectrum of the Hamiltonian, the spectrum contained in Eq.~\eqref{eq:general-qm-neel-oscillation-expval} may be very complicated. Despite encoding two-point correlations, our semiclassical theory is not formulated in terms of superposition of quantum states. Therefore, we should not expect it to reproduce all multi-frequency oscillations that follow from those. However, similar to semiclassical paths within space-time formulations of quantum mechanics [see, for example, Ref.~\citenum{ChaosBook} Chap.~38 or Ref.~\citenum{viscondiSemiclassicalPropagatorGeneralized2015}], we may expect that the semiclassical mapping can recover the mean frequencies $\overline{\omega}_\alpha$ of each correlation component $\alpha$. In this way, our semiclassical theory is closer to experimental protocols such as pump-probe techniques \cite{beaurepaireUltrafastSpinDynamics1996,vankampenAllOpticalProbeCoherent2002,kimelLaserinducedUltrafastSpin2004,kimelUltrafastNonthermalControl2005}, which tend to average over many, possibly inhomogeneous, microscopic subsystems resulting in a mean spectral response.

\subsubsection{Oscillations on a single bond}

As a simple starting point to study the dynamics of spin correlations, we shall begin with an isolated antiferromagnetic dimer on the correlation components. At the semiclassical level, we have the Hamiltonian equations of motion (with bond index surpressed)
\begin{subequations}\label{eq:singlebondheisenbergoperatorequation-semiclassical}
    \begin{align}
\dv{}{t}\mathcal{N}^\pm & = \pm iJ\Big[2\mathcal{N}^z\mathcal{N}^\pm \\ & \qquad\qquad  + \mathcal{N}^z\qty(S(S+1)\hbar^2 - (\mathcal{N}^z)^2)\Big]\notag \\    
\dv{}{t}\mathcal{N}^z & = iJ(\mathcal{N}^+ - \mathcal{N}^-)
\end{align}
\end{subequations}
where we set the magnetization $\mathcal{M}^z=0$ which is constant in time. The original Heisenberg equations for the quantum operators have already been presented in Eqs.~\eqref{eq:singlebondheisenbergoperatorequation}. Together with the Néel-state initial conditions $\mathcal{N}^z=S\hbar$ and $\mathcal{N}^\pm=0$, the dynamical system Eqs.~\eqref{eq:singlebondheisenbergoperatorequation-semiclassical} can be solved analytically [see App.~\ref{App:solution-neel-osc-sc}] and its solution can be expressed in terms of the real components $\mathcal{N}^1 = (\mathcal{N}^+ + \mathcal{N}^-)/2$ and $\mathcal{N}^2 = (\mathcal{N}^+ - \mathcal{N}^-)/(2i)$ as 
\begin{widetext}
    \begin{subequations}\label{eq:solution-singlebond}
    \begin{align}
        & \hspace{-16pt}\text{(semi classical)} & & \hspace{-24pt} \text{(quantum)}\notag\\
        \mathcal{N}^1 & = \frac{S^2\hbar^2}{4}\qty[ \cos(2\omega_\mathrm{sc}t)-1] & \big\langle{\hat{N}^1} \big\rangle & = \Tilde{N}^1_0 + \sum\limits_{m=1}^{2S} \Tilde{N}^1_{2m+1} \cos\qty((2m+1)\omega_0t)\\
            \mathcal{N}^2 & = S\sqrt{\frac{S}{2}}\hbar^2 \sin(\omega_\mathrm{sc}t) & \big\langle{\hat{N}^2} \big\rangle & =  \sum\limits_{m=1}^{2S} \Tilde{N}^2_{m} \sin(m\omega_0t)\\
    \mathcal{N}^z & = S\hbar \cos(\omega_\mathrm{sc}t) & \big\langle{\hat{N}^z} \big\rangle & =  \sum\limits_{m=1}^{2S} \Tilde{N}^z_{m} \cos(m\omega_0t).
    \end{align}
\end{subequations}
\end{widetext}
with the semiclassical oscillation frequency $\omega_\mathrm{sc}=\sqrt{2S}J\hbar$. The quantum evolution of the Néel expectation values $\langle \hat{N}^\alpha\rangle$ in Eqs.~\eqref{eq:solution-singlebond} can also be expressed analytically [see App.~\ref{App:solution-neel-osc-qm} for derivation] and features higher harmonics of the base frequency $\omega_0=J\hbar$ with Fourier coefficients $\Tilde{N}^i_n$ that, in the case of a dimer, are determined by simple Clebsch-Gordon coefficients and include a negative offset $\Tilde{N}^1_0<0$ for $\langle\hat{N}^1\rangle$. We will discuss their concrete distribution shortly. As expected, each semiclassical correlation component oscillates only with a single frequency $\omega_\alpha$ while the exact quantum evolution thereof exhibits a whole frequency spectrum related to the superposition of different energy eigenstates. Note that the component $\mathcal{N}^1$ oscillates at twice the frequency as $\mathcal{N}^2$ and $\mathcal{N}^z$ ($\omega_1 = 2\omega_{2,z}$), which stems from the nonlinearity of the equations of motion and is reminiscent of the parametric resonance known from parallel pumping in nonlinear ferromagnetic resonance experiments \cite{bloembergenRelaxationEffectsFerromagnetic1952,suhlTheoryFerromagneticResonance1957,schlomannRecentDevelopmentsFerromagnetic1960}.

\begin{figure}[h!]
    \centering
    \includegraphics{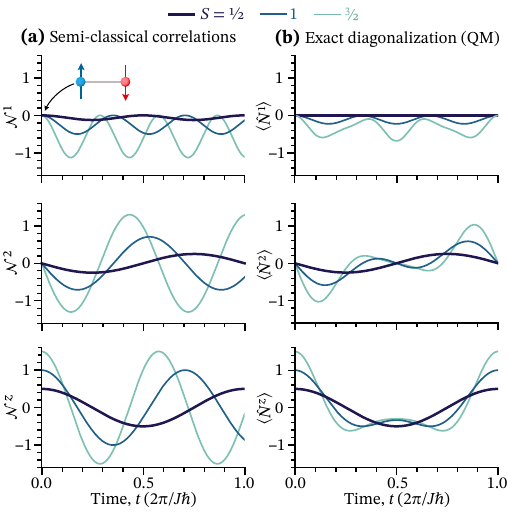}
    \caption{Time evolution of the spin-correlation components $\hat{N}^\alpha$ of an antiferromagnetically coupled dimer with different spin $S$ and initialized in the Néel state (with $\hbar=1$). The curves have been calculated according to \textbf{(a)} the semiclassical theory based on spin-correlations developed in this work and \textbf{(b)} exact diagonalization of the quantum Hamiltonian. In both cases, the magnetization $\langle \hat{M}^z\rangle=\mathcal{M}^z=0$ at all times. }
    \label{fig:neel-osc-single-time-dep}
\end{figure}

The number of harmonics in the quantum evolution is always $2S$ for the case of a single bond. Therefore, at $S=\sfrac{1}{2}$, the quantum evolution of each component is also monochromatic. Indeed, in this case,
\begin{subequations}\label{eq:solution-singlebond_s0p5}
    \begin{align}
        & \hspace{-16pt}\text{(semi classical $S=\sfrac{1}{2}$)} & & \hspace{-24pt}\text{(quantum $S=\sfrac{1}{2}$)}\notag\\
        \mathcal{N}^1 & = \hbar^2\qty[ \cos(2\omega_{0}t)-1] & \expval{\hat{N}^1} & = 0\\
            \mathcal{N}^2 & = \frac{\hbar^2}{4} \sin(\omega_{0} t) & \expval{\hat{N}^2} & =  \frac{\hbar^2}{4} \sin(\omega_{0} t)\\
    \mathcal{N}^z & = \frac{\hbar}{2} \cos(\omega_0 t) & \expval{\hat{N}^z} & =  \frac{\hbar}{2} \cos(\omega_0 t)
    \end{align}
\end{subequations}
which shows that our semiclassical theory not only captures nonclassical dynamics in the spin correlations on dimers, but, for $S=\sfrac{1}{2}$, reproduces the quantum evolution in all components \textit{exactly} except for $\langle \hat{N}^1\rangle$. The solutions in both cases for $S=\sfrac{1}{2}$, 1 and $\sfrac{3}{2}$ are shown in Fig.~\ref{fig:neel-osc-single-time-dep}. While for higher spin $S > \sfrac{1}{2}$, the frequency in the semiclassical limit increases with $\sqrt{2S}$, the quantum expectation values are markedly distorted by the presence of higher harmonics whose number increases with $S$.

\begin{figure}
    \centering
    \includegraphics{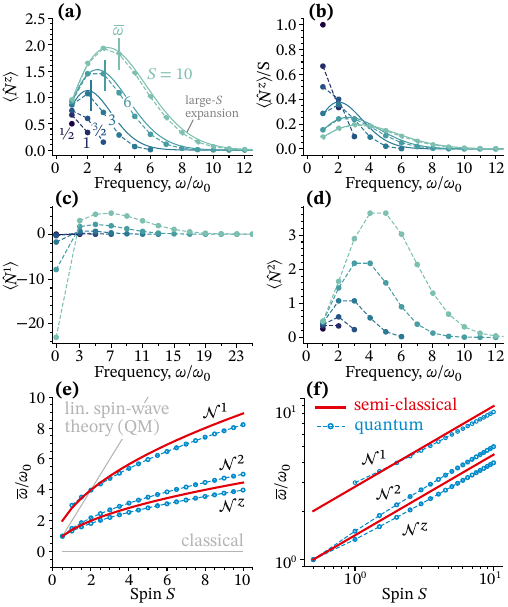} 
    \caption{Comparison of the semiclassical theory with the analytical solution of the quantum Hamiltonian, for the oscillations observed when initializing a single antiferromagnetically coupled dimer initialized in the Néel state. \textbf{(a-d)} Fourier coefficients of the expectation values of $\hat{N}^\alpha$ ($\alpha=1,2,z$) comprising spectra that become more Gaussian-like with increasing $S$. Dashed lines have been added to guide the eye. Solid lines in \textbf{(a)} show the spectrum obtain in the large-$S$ approximation [App.~\ref{App:large-S}]. \textbf{(b)} Fourier spectrum of $\langle \hat{N}^z\rangle$ normalized by $S$. \textbf{(e,f)} Comparison of quantum mean frequencies with the oscillations in the semiclassical limit (on linear and double-logarithmic scale). Thin solid lines show the expected frequency scaling from quantum linear spin-wave theory and classical magnetism for comparison.}
    \label{fig:neel-osc-single-freq-dep}
\end{figure}

In order to understand how the semiclassical theory captures nonclassical dynamics for larger $S$, let us examine the higher harmonics contained in the quantum evolution more closely. To this end, Fig.~\ref{fig:neel-osc-single-freq-dep}(a-c) shows the Fourier coefficients obtained from the analytical solution of the quantum problem for the different Néel expectation values for selected values between $S=\sfrac{1}{2}$ and $S=10$ [App.~\ref{App:solution-neel-osc-qm}]. Although the spectra are discrete, lines have been added to guide the eye. With increasing $S$, the spectra become progressively Gaussian-like, which can be seen by expressing the energy eigenstates using Clebsch-Gordon coefficients and performing a large-$S$ approximation [see App.~\ref{App:large-S}]. As $S$ increases, the maxima of the spectra for the different Néel components are shifted to larger and larger harmonics of the base frequency $\omega_0=J\hbar$. In order to compare this scaling behavior of the semiclassical dynamics, we obtain the mean frequencies $\overline{\omega}_\alpha(S)$ (with $\alpha=1,2,z$) for each component at different values of $S$ of the quantum expectation values (while excluding the large negative constant offset in the $N_n^1$). Figure~\ref{fig:neel-osc-single-freq-dep}(e,f) shows (on linear and double-logarithmic scale) how these mean frequencies exhibit a similar $\sqrt{2S}$-behavior and are well approximated by the semiclassical dynamics in all three correlation components. Indeed, in the large-$S$ approximation [App.~\ref{App:large-S}], one finds that the mean frequencies of the Néel-vector $z$-component go as 
\begin{equation}
    \overline{\omega}_z(S) \approx \sqrt{\pi S/2} \approx 0.89 \sqrt{2S} \approx \omega_\mathrm{sc}(S).
\end{equation} 
It is important to note that the amplitude scaling of the oscillations in $\langle \hat{N}^z\rangle$ normalized by $S$ [see Fig.~\ref{fig:neel-osc-single-freq-dep}(d)] demonstrates how the spectra of these Néel-state oscillations wash out with increasing $S$ and become flat in the classical limit. This stresses again that these oscillations are absent in classical magnetism. Moreover, quantum linear spin-wave theory predicts a frequency scaling proportional to $2S$ instead of $\sqrt{2S}$ \cite{auerbachInteractingElectronsQuantum1994}. This means that the dynamics captured by our semiclassical spin-correlation theory not only capture nonclassical effects but also nonlinear effects. Such effects nonlinear in $S$ are expected since, in contrast to weakly nonlinear expansions in terms of bosons, we have not made any approximations to the commutation relations of the spin operators before mapping them onto semiclassical correlations. Moreover, our theory suggests a spin-correlation analogue of the nonlinear parametric resonance present in classical magnetism, where one component ($\hat{N}^1$ in the present case) oscillates at twice the frequency due to the nonlinear terms emerging from the $\mathfrak{su}(2)$ algebra. Importantly, the exact unitary quantum evolution also resembles this frequency doubling [as visible in Fig.~\ref{fig:neel-osc-single-freq-dep}(f)] for $S>1/2$.

\subsubsection{Scaling with system size}

\begin{figure*}
    \centering
    \includegraphics{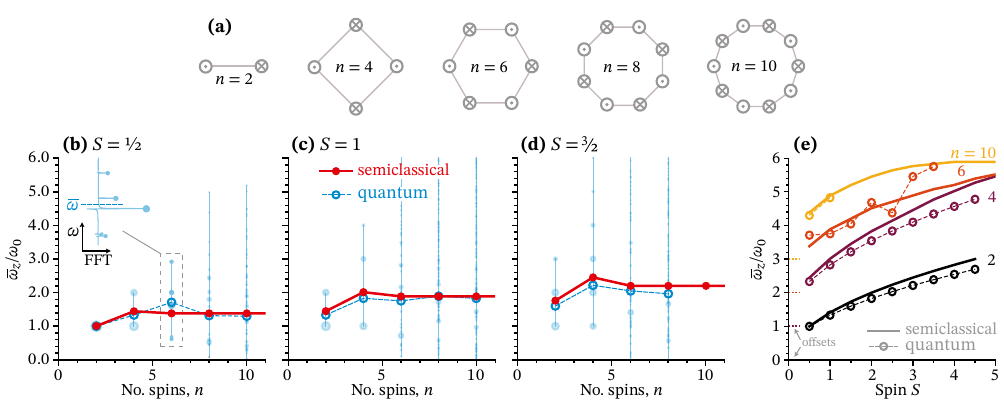}
        \caption{Comparison of the oscillation frequencies obtained with exact diagonalization and the semiclassical mapping as a function of the number of spins $n$ in \textbf{(a)} different one-dimensional antiferromagnetic chains with periodic boundary conditions and initialized in a Néel state. \textbf{(b-d)} (Mean) frequency scaling of the Néel vector $z$-component in both cases for $S=\sfrac{1}{2}$, 1 and $\sfrac{3}{2}$. The area of the faded blue dots in the background represent the original Fourier coefficients in ED of which the mean is taken for each $n$ [see the inset in \textbf{(b)}]. \textbf{(e)} Same comparison up to even larger spin $S$. For better visiblity, the curves for $n=4,8$ and $10$ have been offset by increasing multiples of $\omega_0$.}
    \label{fig:neel-osc-scaling}
\end{figure*}

Next, let us explore the behavior of our semiclassical theory with increasing system size. For this, we will focus on the Néel-state oscillations in one-dimensional spin chains with $n\geq 2$ spins and periodic boundary conditions [Fig.~\ref{fig:neel-osc-scaling}(a)], for which we can still perform full diagonalization for small spin quantum number $S$. Here, we performed diagonalization of the Hamiltonian within the antiferromagnetic sector ($S_\mathrm{tot}=0$) and the unitary evolution of the Néel state using the \textsc{QuSpin} package  \cite{weinbergQuSpinPythonPackage2017,weinbergQuSpinPythonPackage2019}. Figure~\ref{fig:neel-osc-scaling}(b) shows the (mean) frequencies for $S=\sfrac{1}{2}$ of $\langle \hat{N}^z\rangle$ and $\mathcal{N}^z$ according to ED and our semiclassical theory, respectively. Due to the translational invariance of both the Hamiltonian and (in our sign structure) the initial state, we only show the frequencies on a single bond. From our simulations, we have confirmed that the time evolution is indeed identical on all bonds. As with increasing $S$, the number of frequency components in the exact quantum evolution increases with the system size. However, as the inset in Fig.~\ref{fig:neel-osc-scaling}(a) shows, for $n\geq 6$, the Fourier spectrum of $\langle \hat{N}^z \rangle$ is no longer comprised of higher harmonics of the base frequency $\omega_0$ only, but instead features an abundance of other modes. Throughout Figs.~\ref{fig:neel-osc-scaling}(b-d), these additional modes are shown for each number of spins $n$ as faded dots, whose area is proportional to the spectral weight of the corresponding peak in the Fourier spectrum of $\langle{\hat{N}^z}\rangle$. Despite these additional spectral components, the scaling of the corresponding mean frequency $\overline{\omega}_z$ in the system size is still well-approximated by our semiclassical theory, as shown in Figs.~\ref{fig:neel-osc-scaling}(b-c) for $S=\sfrac{1}{2}, 1$ and $\sfrac{3}{2}$. Additionally, Fig.~\ref{fig:neel-osc-scaling}(e) shows the scaling behavior for different $n\geq 4$ as a function of the spin $S$, showing a similar $\sqrt{2S}$ behavior as in the case of a single dimer ($n=2$) before in Fig.~\ref{fig:neel-osc-single-freq-dep} and a remarkable agreement between ED and the semiclassical dynamics for $n=4$. %Additionally, Fig.~\ref{fig:neel-osc-scaling}(e) shows the system-size scaling of the oscillation frequency $\overline{\omega}_z$ for even larger $S$. While the oscillation frequency converges to an $S$-dependent value, the system size at which this convergence occurs seems to increase with $S$. 

We summarize that the semiclassical spin-correlation theory captures coherent oscillations of the Néel vector in Heisenberg antiferromagnets at a frequency that approximates the mean frequency of the exact quantum evolution, when the system is initialized in the antiferromagnetic Néel state. This is in stark contrast to classical magnetism, in which the Néel state is an equilibrium state that does not exhibit any dynamics. Our theory predicts a $\sqrt{2S}$ scaling of the oscillation mean frequencies, which is clearly different from the typical $2S$-scaling found in linear spin-wave theory. Despite being built only on two-point correlations, our semiclassical theory still approximates the mean frequencies obtained by exact diagonalization well, even beyond the isolated dimer. This demonstrates that semiclassical correlations can capture collective nonlinear dynamics that are completely nonclassical. 

%\section{Discusssion}

\section{Conclusions and Outlook}\label{sec:conclusion_outlook}

We have developed a semiclassical theory of spin–correlation dynamics that extends the conceptual foundation of classical nonlinear magnetism to incorporate nonclassical dynamics. However, rather than mapping quantum spin operators directly to classical spin vectors, our approach establishes a correspondence at the level of bondwise spin correlations, preserving the structure of the Heisenberg equations of motion while retaining the quantum spin Casimir $S(S+1)\hbar^2$, which reflects the inclusion of zero-point fluctuations. Starting from the exact $\mathfrak{su}(2)$ commutation relations, this mapping yields fully nonlinear and closed Hamiltonian equations for spin correlations, expressed in terms of Poisson brackets. 

By incorporating zero-point fluctuations at the level of two-spin, rather than single-spin, observables, our semiclassical framework achieves two key advances. (i) It captures essential features of dynamics that are simultaneously nonlinear and nonclassical, such as the quantum oscillations of the antiferromagnetic Néel state. In this case, our theory predicts an oscillation frequency scaling $\omega\propto \sqrt{2S}$, which closely matches the mean oscillation frequencies obtained from exact diagonalization. This behavior is neither reproduced by classical magnetism -- where such oscillations are entirely absent -- nor by quantum linear spin-wave theory. (ii) Our approach restores a geometric intuition for nonlinear quantum spin dynamics. Much like classical spins, semiclassical correlations evolve on a curved phase space, giving rise to nonlinear terms in the equations of motion analogous to the cross-product structure of the Landau–Lifshitz equation. As a result, semiclassical spin correlations exhibit nonlinear dynamical signatures, such as components oscillating at twice the fundamental frequency, directly reminiscent of parametric resonance in classical magnetism that is understood by considering the trajectory of classical spins being constrained to a sphere. Although these features also appear in the exact unitary dynamics, they are far less transparent in full quantum theory; in contrast, our semiclassical formulation makes their physical origin more explicit.

A distinctive feature of semiclassical spin correlations is their nonlocality: While classical spins obey a local length constraint, semiclassical correlations are connected across the lattice through a global constraint arising from their operator algebra. Consequently, correlations between any two sites constrain each other, reflecting a distributed conservation of spin magnitude and leading to $\mathcal{O}(n^2)$ equations for $n$ spins. Conceptually, this connects our theory to bond-operator methods, Schwinger-boson formalisms, and valence-bond approaches, but within a fully nonlinear semiclassical framework that remains directly formulated in terms of observables. Importantly, this nonlocal constraint structure of spin correlations was absent in previous formulations relying on momentum-space correlations within linear spin-wave theory, where correlations associated with distinct wave vectors commute with each other \cite{bossiniLaserdrivenQuantumMagnonics2019,formisanoCoherentTHzSpin2024a}.

Lastly, due to the geometric analogies with classical magnetism, our theory allows for the consistent inclusion of phenomenological at the level of correlations, as relaxation on the semiclassical phase space. This provides a natural semiclassical analogue of the Landau–Lifshitz–Gilbert equation for spin correlations, offering a compact and physically transparent way to describe relaxation and preparation of equilibrium states, without invoking the full machinery of Lindblad or density-matrix dynamics, that are required in full quantum methods.

Although the present work focuses on the Heisenberg Hamiltonian, the framework can be readily extended to include magnetic anisotropies, Dzyaloshinskii–Moriya interactions, external fields, and coupling to particle baths. An open question is whether one can define an effective magnetic field and torque for spin correlations that parallels the classical picture, thereby completing the geometric intuition of nonlinear magnetism in the semiclassical domain. 

By bridging classical nonlinear and quantum magnetism through describing the dynamics of semiclassical spin correlations, our theory provides a new conceptual foundation for studying nonlinear quantum spin phenomena in magnetic materials.

\iffalse

\begin{itemize}
    \item TODO mention inability to capture two-magnon excitations
    \item TODO outlook quench of XXZ model
    \item TODO works for any S!
    \item embraces full nonlinearity while avoiding short-comings of bosonization
    \item compare to schwinger boson and bond operator methods
    \item bridge to VBS
    \item TODO compare with truncated Wigner transform
    \item reformulates magnetism on level of correlations without linearization.
    \item instead of trying to solve the quantum problem more efficiently we go the other route, by starting with classical magnetism and trying to extend it to capture certain quantum effects 
    \item Johan: "makes it very easy to study dissipation, which otherwise requires much more complicated Lindblad dynamics etc, Hence, through the lens of classical nonlinear magnetism, certain quantum features become much more easy to deal with, without having to learn the much more involved and complicated existing quantum descriptions. This is something often overlooked by quantum theorists and a key contribution of the paper."
\end{itemize}

\fi 

\begin{acknowledgments}
We gratefully acknowledge support and useful comments on the manuscript by Hrvoje Vrcan and fruitful discussions with Rein Liefferink and Alexander Mook. L.K. gratefully acknowledges funding by the Radboud Excellence Initiative. Furthermore, funding by the Dutch Research Council (NWO) via VIDI project no. 223.157 (CHASEMAG) and by the European Union via Horizon Europe project no. 101070290 (NIMFEIA) is gratefully acknowledged.
\end{acknowledgments}

\appendix

\section{Exact Néel evolution for two spin-S particles}

Here, we derive the time evolution of the correlation components of single antiferromagnetically coupled ($J>0$) bond with Hamiltonian
\begin{align}
      \hat{H} & = J\hat{\boldsymbol{S}}_1 \cdot \hat{\boldsymbol{S}}_2 \label{eq:hamiltonianJS1S2} \\ 
      \rightsquigarrow \quad \mathcal{H}_\mathrm{sc}& =J\qty[\mathcal{N}^+ + \mathcal{N}^- + \qty(\mathcal{M}^z)^2 - \qty(\mathcal{N}^z)^2] \label{eq:App-hamiltonian-sc}
\end{align}
initalized in the Néel state, in the semiclassical limit and according to quantum mechanics.

\subsection{Semiclassical dynamics}\label{App:solution-neel-osc-sc}

In the semiclassical theory one has $\mathcal{N}^z=S\hbar$ and $\mathcal{M}^z = \mathcal{N}^\pm =0$ in the initial Néel state (at $t=0$).
Since $\mathcal{M}^z$ is time-independent in the case of a single bond, we have the equations of motion
\begin{align}
   \dv{}{t}\mathcal{N}^\pm & = \pm iJ\mathcal{N}^z\qty[2\mathcal{N}^\pm + S(S+1)\hbar^2 - (\mathcal{N}^z)^2]\label{eq:npnmdec}\\ 
   \dv{}{t}\mathcal{N}^z & = iJ(\mathcal{N}^+ - \mathcal{N}^-)
\end{align}
As the energy $\mathcal{H}_\mathrm{sc}/J=-S^2\hbar^2 = \mathcal{N}^+ + \mathcal{N}^- - \qty(\mathcal{N}^z)^2$ is conserved, we may rewrite Eq.~\eqref{eq:npnmdec} as 
\begin{align}
   \dv{}{t}\mathcal{N}^\pm & =  iJ\mathcal{N}^z\qty[\mathcal{N}^+ - \mathcal{N}^- \pm S^2\hbar^2 ]
\end{align}
Introducing $\mathcal{N}^1 = \tfrac{1}{2}(\mathcal{N}^+ + \mathcal{N}^-)$ and $\mathcal{N}^2 = \tfrac{1}{2i}(\mathcal{N}^+ - \mathcal{N}^-)$ leads to the coupled system
\begin{align}
      \dv{}{t}\mathcal{N}^1 & = -2J\mathcal{N}^z\mathcal{N}^2 \label{eq:N1eqApp} \\
      \dv{}{t}\mathcal{N}^2 & = JS\hbar^2\mathcal{N}^z \label{eq:N2eqApp}\\
   \dv{}{t}\mathcal{N}^z & = -2J\mathcal{N}^2\label{eq:NzeqApp}
\end{align}
Taking another time derivative of Eqs.~\eqref{eq:N2eqApp} and \eqref{eq:NzeqApp}
and inserting them into each other yields
\begin{equation}
    \dv[2]{}{t}\mathcal{N}^{2,z} = - \omega_\mathrm{sc}^2 \mathcal{N}^{2,z} \qquad \text{with} \qquad \omega_\mathrm{sc} = \sqrt{2S}J\hbar.
\end{equation}
From the aforementioned initial conditions in conjunction with Eqs.~\eqref{eq:N2eqApp} and \eqref{eq:NzeqApp} it also follows that $\mathrm{d}\mathcal{N}^2/\mathrm{d}t = J\hbar^3 S^2$ and $\mathrm{d}\mathcal{N}^z/\mathrm{d}t = 0$ at $t=0$ and therefore
\begin{align}
    \mathcal{N}^2 & = S\sqrt{\frac{S}{2}}\hbar^2 \sin(\omega_\mathrm{sc}t)\\
    \mathcal{N}^z & = S\hbar \cos(\omega_\mathrm{sc}t).
\end{align}
The missing correlation component can either be obtained by integrating the right-hand side of Eq.~\eqref{eq:N1eqApp} or simply again using the energy conservation
\begin{equation}
    \mathcal{H}_\mathrm{sc} = -JS^2\hbar^2 = J\qty[2 \mathcal{N}^1  - \qty(\mathcal{N}^z)^2]
\end{equation}
which results in 
\begin{align}
    \mathcal{N}^1 & = -\frac{S^2\hbar^2}{2}\sin^2(\omega_\mathrm{sc}t) = \frac{S^2\hbar^2}{4}\qty[ \cos(2\omega_\mathrm{sc}t)-1].
\end{align}
This resembles a well-known feature of nonlinear single-spin dynamics, where the local constraint also leads to frequency doubling.

\subsection{Quantum solution}\label{App:solution-neel-osc-qm}

%\begin{itemize}
  %  \item \textbf{LK: Please use the same formatting and nomenclature as in the main text. e.g. bold vectors, name of Néel state, etc.}
    %\item \textbf{LK (done): Rename J to Stot to avoid confusion with exchange constant? -> renamed J to L.}
    %\item \textbf{LK (done): Throughout the manuscript we do not set J (exchange) to 1 in the equations. Maybe also better not to do it here.}
 %   \item \textbf{PC: I got rid of the shorthand notation for CG coefficients as it did not save space (it appears only where we need 2 lines anyway)}
%\end{itemize}
 
In this section, we derive the analytical unitary time evolution of two antiferromagnetically coupled spin-$S$ particles initialized in the Néel state, with the Hamiltonian given in Eq.~\eqref{eq:hamiltonianJS1S2}. As is well-known from angular momentum theory, this Hamiltonian is completely diagonalized by the eigenstates of ${\hat{\boldsymbol{S}}_\mathrm{tot}}^2$ and $\hat{S}_\mathrm{tot}^z$ (with $\hat{\boldsymbol{S}}_\mathrm{tot} = \hat{\boldsymbol{S}}_1 + \hat{\boldsymbol{S}}_2$). These two operators commute with the Hamiltonian and with each other. Therefore, the total spin quantum number $L = 0,1,...,2S$ and magnetic quantum number $M = -L,-L+1,...,L$ are suitable quantum numbers to label the eigenstates $\ket{LM}$. The energies of the eigenstates are
\begin{equation}
    E(L) = J \hbar^2 \left[ \tfrac{1}{2} L(L + 1) - S(S+1) \right]
\end{equation}
with each energy level being $(2L+1)$ times degenerate, once for each possible value of $M$. 

Now, we wish to express the Néel state in terms of these eigenstates such that we can easily apply the time evolution operator. The Néel state is defined as the fully polarized state in the uncoupled basis 
\begin{equation}
    \ket{\psi(0)} = \ket{S_1 = S, m_1 = S, S_2 = S, m_2 = -S}.
\end{equation}
Note that since the Néel state has $M=0$ and $M$ is a good quantum number, only the eigenstates with $M=0$ will have nonzero overlap with the Néel state. Under time-evolution, we find
\begin{equation} \label{eq:neel_t_wavefunction}
    \ket{\psi(t)} = \sum_{L=0}^{2S} C_{SSS(-S)}^{L0} e^{-iE_{L}t/\hbar} \ket{L,0},
\end{equation}
with $C_{SSS(-S)}^{L0}$ being the Clebsch-Gordan coefficients that are known exactly as
\begin{equation} \label{eq:exact_CG} 
    C_{SSS(-S)}^{L0} = (2S)!\sqrt{\frac{2L+1}{(L+2S+1)!(2S-L)!}}.
\end{equation}
% could cite \cite{varshalovichQuantumTheoryAngular1988} here?
We now proceed to calculate the expectation value of the Néel operator:
\begin{equation} \label{eq:neel_expectation_value_braket}
    \langle \hat{N}^z(t) \rangle = \bra{\psi(t)} \frac{\hat{S}_1^z - \hat{S}_2^z}{2} \ket{\psi(t)}
\end{equation}
%
%Decide if we want to write this term out further
Expanding $\ket{\psi(t)}$ as in \eqref{eq:neel_t_wavefunction}, we note that all components of this expression are known, except for the matrix element $\bra{L'0}\hat{S}_1^z - \hat{S}_2^z \ket{L 0}$.
Since all magnetizations in the matrix element are zero, we can deduce $\bra{L'0}\hat{S}_1^z - \hat{S}_2^z \ket{L 0} = 2\bra{L'0} \hat{S}_1^z \ket{L 0}$. Using the expression for matrix elements of spherical tensor operators that depend on a single subsystem from Ref.~\citenum{varshalovichQuantumTheoryAngular1988}, Sec. 13.1.5, Eq.~(40), we obtain
\begin{equation} \label{eq:matrix_element_Nz}
    \begin{split}
        \bra{L' 0} \hat{S}_1^z \ket{L 0} = &(-1)^{L+2S-1} \sqrt{(2L+1)} \langle L' 0 | L 0 1 0 \rangle \\
        & \times \begin{Bmatrix} L' 1 L \\ S S S \end{Bmatrix} \hbar \sqrt{S(S+1)(2S+1)}
    \end{split}
\end{equation}
with $\{ L' 1 L ; S S S \}$ being the Wigner 6j symbol.
From triangle relations and symmetry properties of the Clebsch-Gordan coefficient, we see that $\bra{L'0}S_1^z\ket{L0}$ is nonzero only if $L$ and $L'$ differ by exactly one, which greatly simplifies the final expression.  
Using this fact, along with $E_{L+1}-E_L = J\hbar^2(L+1)$, we obtain the final exact expression
%
%\begin{equation} \label{eq:final_exact_Nz(t)}
%    \begin{split}
%        \langle \hat{N}^z(t) \rangle & = 4 \sum_{n=1}^{2S+1} \langle S S S (-S) | (n-1) 0 \rangle \langle S S S (-S)|n 0 \rangle \\ & \qquad  \times \bra{(n-1) 0}\hat{S}_1^z \ket{n 0} \cos(nJt)
%    \end{split}
%\end{equation}
%
%
\begin{equation} \label{eq:final_exact_Nz(t)}
    \begin{split}
        \langle \hat{N}^z(t) \rangle & = \sum_{n=1}^{2S+1} \tilde{N}^z_n \cos(nJ\hbar t)
    \end{split}
\end{equation}
with
\begin{equation}
    \tilde{N}^z_n = 2 C_{SSS(-S)}^{(n-1)0} C_{SSS(-S)}^{n0} \bra{(n-1) 0}\hat{S}_1^z \ket{n 0}
\end{equation}
with the matrix element $\bra{(n-1) 0}S_1^z \ket{n 0}$ given by Eq.~\eqref{eq:matrix_element_Nz}.

Now we proceed to obtain the expressions for $\langle \hat{N}^1(t) \rangle$ and $\langle \hat{N}^2(t) \rangle$ as defined in the main text. Following similar arguments as for $\langle \hat{N}^z(t)\rangle$, we have
\begin{equation}
    \begin{split}
        & \bra{L' 0}\hat{S}_1^\pm \hat{S}_2^\mp \ket{L 0}  = \hbar^2\sqrt{2L+1} S(S+1)(2S+1) \\ 
        & \qquad \times\sum_{c\gamma}\sqrt{2c+1}C^{L'0}_{L0c\gamma} C^{c\gamma}_{1,\pm1,1,\mp1}\begin{Bmatrix} 1 1 c \\ S S L' \\ S S L \end{Bmatrix}
    \end{split}
\end{equation}
with $\{11c; SSL^\prime; SSL\}$ being a Wigner 9j-symbol. It is easy to see that $c=0,1,2$ and $\gamma=0$.
Then for the matrix elements of $\hat{N}^1$ and $\hat{N}^2$, we get
\begin{widetext}
    \begin{equation}
    \begin{split}
        \bra{L'0}\hat{N}^1\ket{L0}  = \frac{\hbar^2}{2}\sqrt{2L+1} S(S+1)(2S+1) \Biggl( C^{L'0}_{L000}C^{00}_{1,1,1,-1}\begin{Bmatrix} 1 1 0 \\ S S L' \\ S S L \end{Bmatrix} 
         + \sqrt{5}C^{L'0}_{L020} C^{20}_{1,-1,1,+1}\begin{Bmatrix} 1 1 2 \\ S S L' \\ S S L \end{Bmatrix} \Biggr)
    \end{split}
\end{equation}
and
\begin{equation}
    \bra{L'0}\hat{N}^2\ket{L0} = \frac{\hbar^2}{2i}\sqrt{2L+1} S(S+1)(2S+1) \sqrt{3} \ C^{L'0}_{L010}C^{10}_{1,1,1,-1} \begin{Bmatrix} 1 1 1 \\ S S L' \\ S S L \end{Bmatrix}
\end{equation}
\end{widetext}
From triangle relations and symmetries of the Clebsch-Gordan coefficients and 9j-symbols, it follows that $\bra{L'0}\hat{N}^1\ket{L0}$ can only be nonzero if $L$ and $L'$ differ by exactly zero or two. Similarly, it follows that $\bra{L'0}\hat{N}^2\ket{L0}$ can only be nonzero if $L$ and $L'$ differ by exactly one.

After some manipulations, we finally have
\begin{equation}
    \langle \hat{N}^1(t) \rangle = \tilde{N}_0^1 + \sum_{n=1}^{2S+1} \tilde{N}^1_n \cos((2n+1)J\hbar t)
\end{equation}
with
\begin{equation}
    \tilde{N}^1_n = 2 C_{SSS(-S)}^{(n-1)0} C_{SSS(-S)}^{(n+1)0} \bra{(n+1) 0} \hat{N}^1 \ket{(n-1) 0}
\end{equation}
and
\begin{equation}
    \tilde{N^1_0} = \sum_{L=0}^{2S} \left( C_{SSS(-S)}^{L0} \right)^2 \bra{L 0} \hat{N}^1 \ket{L 0}.
\end{equation}
For $\langle \hat{N}^2(t) \rangle$ we get
\begin{equation}
    \langle \hat{N}^2(t) \rangle = \sum_{n=1}^{2S+1} \tilde{N}^2_n \sin(nJ\hbar t)
\end{equation}
with
\begin{equation}
    \tilde{N}^2_n = 2 C_{SSS(-S)}^{{(n-1)0}} C_{SSS(-S)}^{n0} \left| \bra{n 0} \hat{N}^2 \ket{(n-1) 0} \right| 
\end{equation}

\section{Asymptotics of quantum Néel-state oscillations in large-$S$ limit}\label{App:large-S} 
 
To extract the scaling of the mean frequency of $\langle \hat{N}^z(t) \rangle$ from Eq.~\eqref{eq:final_exact_Nz(t)}, we examine the asymptotics of this expression in the large-$S$ limit. Let us start by obtaining the asymptotics of the CG coefficients \eqref{eq:exact_CG}, which turn out to be very well approximated by a Gaussian. To see this, we start by performing a Stirling approximation for the factorials, $n! \approx \sqrt{2\pi n}\left( \frac{n}{e} \right)^n$. We arrive at the expression
\begin{equation}
    \begin{split}
         C_{SSS(-S)}^{L0} &\approx \sqrt{e} \sqrt{\frac{(2S)(2L+1)}{\sqrt{L+2S+1}\sqrt{2S-L}}} \\
         &\times \frac{(2S)^{2S}}{\sqrt{(2S+L+1)^{2S+L+1} (2S-L)^{2S-L}}}.
    \end{split}
\end{equation}
We further approximate the fraction with the $n^n$ terms by taking its logarithm and performing a Taylor expansion of each logarithm term around $\ln(2S)$. Simplifying this and absorbing $\sqrt{e}$ into the exponential, we find
\begin{equation}
C_{SSS(-S)}^{L0} \approx\sqrt{\frac{2L+1}{\sqrt{L+2S+1}\sqrt{2S-L}}} \ e^{Q_S(L)}
\end{equation} 
with
\begin{equation}
    Q_S(L) = \left( \frac{3}{16S^2} - \frac{1}{4S} \right) \left( L^2 + L \right) + \left( \frac{1}{16S^2} - \frac{1}{8S} \right).
\end{equation}
Note the Gaussian form of this expression.

We now proceed by performing the asymptotics for the 6j symbol. We use Eq.~(4) in Sec.~9.9.1 of Ref.~\citenum{varshalovichQuantumTheoryAngular1988}
\begin{equation}
     \begin{Bmatrix} L' 1 L \\ S S S \end{Bmatrix} \approx \frac{(-1)^L}{\sqrt{2S}\sqrt{2L+1}} C_{L'010}^{L0} 
\end{equation}
%

%Something should still be said about the "two square roots" part of the equation, which we approximate as a linear function n/S.

Combining these approximations, we obtain an expression for $\tilde{N}^z_n$ which consists of a Gaussian with a rather involved prefactor which depends on $n$. We found that the prefactor can be very well approximated by a linear function $2n$. This approximation becomes inaccurate when $n$ gets large, but this does not matter as in this regime the amplitudes are suppressed exponentially strongly by the Gaussian.

We then arrive at the final asymptotic expression for the Fourier coefficient $\tilde{N}^z_n$ which corresponds to the frequency $nJ\hbar$:
\begin{equation}
     \tilde{N}^z_n \approx 2n \cdot \exp\left[ \left( \frac{3}{8S^2} - \frac{1}{2S} \right) n^2 + \left( \frac{1}{8S^2} - \frac{1}{4S} \right) \right]
\end{equation}
We then regard this as a continuous distribution and calculate its mean frequency
\begin{equation}
     f_{mean} = J\hbar\sqrt{\frac{2\pi S^2}{4S-3}} \approx J\hbar\sqrt{\frac{\pi S}{2}}.
\end{equation}

\section{Connection to linear correlation theory and emergent SU(1,1)}\label{sec:su11}

Here we briefly show how the Néel correlation operators $\hat{N}^\alpha_\nu$ on a bond reduce to a set of Perelomov operators admitting a $\mathfrak{su}(1,1)$ algebra, when linearized around the antiferromagnetic Néel state. These operators are directly related to two-magnon excitations. The spin operators can be expressed in linear spin-wave theory by using a Holstein-Primakoff expansion as 
\begin{align}
    \hat{S}^z_i & = \hbar(S-\hat{a}_i^\dagger \hat{a}_i) &     \hat{S}^z_j & = - \hbar(S-\hat{b}_j^\dagger \hat{b}_j)\\
    \hat{S}^-_i & =   \hbar\sqrt{2S}\hat{a}_i^\dagger &     \hat{S}^-_j & = \hbar\sqrt{2S} \hat{b}_j\\
    \hat{S}^+_i & = \hbar\sqrt{2S}\hat{a}_i &     \hat{S}^+_j & =\hbar\sqrt{2S} \hat{b}_j^\dagger
\end{align}
with $\hat{a}_i^{(\dagger)}$ and $\hat{b}_j^{(\dagger)}$ being the boson annihilation (creation) operators on sublattice $\mathbb{A}$ and $\mathbb{B}$, respectively. In this approximation, the Néel operators become
\begin{align}
    \hat{M}^z & = \frac{\hbar}{2}\qty(-\hat{a}^\dagger_i\hat{a}_i + \hat{b}^\dagger_j\hat{b}_j)&     
        \hat{N}^z & = \frac{\hbar}{2}\qty(2S-\hat{a}^\dagger_i\hat{a}_i - \hat{b}^\dagger_j\hat{b}_j) \\
         \hat{N}^+ & = S\hbar^2 \hat{a}_i\hat{b}_j &  \hat{N}^- & = S\hbar^2 \hat{a}_i^\dagger\hat{b}_j^\dagger.
\end{align}
Using the canonical commutation relations of the boson operators, one can show that all single-bond commutators of the Néel operators in Eqs.~\eqref{eq:single-bond-commutators} remain unchanged under this approximation, except
\begin{align}
         \qty[\hat{N}^+,\hat{N}^-] &= S^2\hbar^4 \qty(\hat{a}_i^\dagger\hat{a}_i + \hat{b}_j^\dagger\hat{b}_j+1)\\
           & = S^2\hbar^4 \qty(\frac{(2S+1)}{2}-\frac{\hat{N}^z}{\hbar})
\end{align}
One can now introduce the boson-pair operators
\begin{align}
  \hat{K}^\pm   & = \frac{\hat{N}^\mp }{S\hbar^2} \qquad\text{and}\qquad  \hat{K}^z   = \frac{S^2\hbar^4}{2} \qty(\frac{(2S+1)}{2}-\frac{\hat{N}^z}{\hbar}).
\end{align}
From the commutation relations of the linearized Néel operators it follows that these new operators are the Perelomov operators, which satisfy the $\mathfrak{su}(1,1)$ algebra 
\begin{equation}
    [\hat{K}^z,\hat{K}^\pm]  = \pm \hat{K}^\pm, \qquad [\hat{K}^+,\hat{K}^-] = -2\hat{K}^z.
\end{equation}
which differs from $\mathfrak{su}(2)$ by the additional minus sign. 

\newpage 

\begin{widetext}
    \section{Collection of multi-bond commutators and brackets}\label{App:commsbracks}

Here, summarize the commutation relations of the Néel operators and the dissipative brackets of the semiclassical correlations. Operators trivially commute between disconnected bonds.

\begin{table*}[h!]
\centering
\begin{tblr}{
  cells = {c},
  hlines,
  vlines,
  hline{1-2,9,16} = {-}{0.08em},
}
& same bond &  connected bonds\\
 $\qty[\hat N^z_{{\nu}}, \hat N^z_{{\mu}}]$ & 0 & 0  \\[.3cm]
         $\qty[\hat N^z_{{\nu}}, \hat M^z_{{\mu}}]$ & 0 & 0  \\[.3cm]
          $\qty[\hat M^z_{{\nu}}, \hat M^z_{{\mu}}]$ & 0 & 0  \\[.3cm]
         $\qty[\hat N^z_{{\nu}}, \hat N^\pm_{{\mu}}]$ & $\pm\hbar \hat N^\pm_{{\mu}}$ & $\pm\frac{\hbar}{2} \hat N^\pm_{{\mu}}$ \\[.3cm]
         $\qty[\hat M^z_{{\nu}}, \hat N^\pm_{{\mu}}]$ & 0 & $\pm\frac{\hbar}{2} \varsigma_{\nu\mu} \hat N^\pm_{{\mu}}$   \\[.3cm]
         $\qty[\hat N^\pm_{{\nu}}, \hat N^\pm_{{\mu}}]$ & 0 & 0  \\[.3cm]
         $\qty[\hat N^+_{{\nu}}, \hat N^-_{{\mu}}]$ & $ \hbar \hat N^z_\mu \qty[S(S+1)\hbar^2 + (\hat M^z_\mu)^2 - (\hat N^z_\mu)^2]$ &  $\varsigma_{\nu\mu}\hbar \hat C^\mp_{\nu,\mu} \hat{S}_j^z $  \\
$\qty{\mathcal N^z_{{\nu}}, \mathcal N^z_{{\mu}}}_+$ & $-\frac{(\mathcal{N}_\nu^z)^2 + (\mathcal{M}_\nu^z)^2}{2} + \frac{S(S+1)\hbar^2}{2}$ & $\frac{S(S+1)\hbar^2}{4} - \frac{\mathcal{S}^z_j}{4}$  \\[.3cm]
         $\qty{\mathcal N^z_{{\nu}}, \mathcal M^z_{{\mu}}}_+$ & $-\mathcal{N}^z_\nu \mathcal{M}_\mu^z$ & $\varsigma_{\nu\mu}\qty[\frac{S(S+1)\hbar^2}{4} - \frac{\mathcal{S}^z_j}{4}]$  \\[.3cm]
          $\qty{\mathcal M^z_{{\nu}}, \mathcal M^z_{{\mu}}}_+$ &  $-\frac{(\mathcal{N}_\nu^z)^2 + (\mathcal{M}_\nu^z)^2}{2} + \frac{S(S+1)\hbar^2}{2}$ & $\frac{S(S+1)\hbar^2}{4} - \frac{\mathcal{S}^z_j}{4}$   \\[.3cm]
         $\qty{\mathcal N^z_{{\nu}}, \mathcal N^\pm_{{\mu}}}_+$ & $-\mathcal{N}^z_\nu \mathcal{N}_\mu^\pm$   &  $-\frac{\varsigma_{\nu\mu}\mathcal{N}_\mu^\pm \mathcal{S}_j^z}{2}$ \\[.3cm]
         $\qty{\mathcal M^z_{{\nu}}, \mathcal N^\pm_{{\mu}}}_+$ & $-\mathcal{M}^z_\nu \mathcal{N}_\mu^\pm $  &   $-\frac{\mathcal{N}_\mu^\pm \mathcal{S}_j^z}{2}$  \\[.3cm]
         $\qty{\mathcal N^\pm_{{\nu}}, \mathcal N^\pm_{{\mu}}}_+$ & $-(\mathcal{N}^\pm_\nu)^2$  & $-\mathcal{N}^\pm_\nu \mathcal{N}^\pm_\mu$ \\[.3cm]
         $\qty{\mathcal N^+_{{\nu}}, \mathcal N^-_{{\mu}}}_+$ & $ - 2\mathcal{N}_\nu^+\mathcal{N}_\nu^- + S(S+1)\hbar^2\qty[S(S+1)\hbar^2 -(\mathcal{N}_\nu^{z})^2 - (\mathcal{M}_\nu^{z})^2] $  &     $ \qquad S(S+1)\hbar^2 \mathcal{C}_{\nu,\mu}^\mp -\mathcal{N}^\pm_\nu \mathcal{N}^\pm_\mu $
\end{tblr}
    \caption{Commutation relations between interlattice correlation operators and dissipative brackets between semiclassical interlattice correlations. Here, $j=\mathrm{shared}(\nu,\mu)$ denotes the shared spin in a connected bond. Moreover, $\varsigma_{\nu\mu} = \pm 1$ and the signs in $\hat{C}^\mp$  and $\mathcal{C}^\mp$ are taken depending on whether $j$ is in $\mathbb{A}$ or $\mathbb{B}$. The spin components $\hat{S}^z_j$ and $\mathcal{S}^z_j$ can be replaced by adjacent correlations using Eq.~\eqref{eq:replace-Sz}.}
    \label{tab:placeholder}
\end{table*}

\newpage

\begin{table*}[h!]
\centering
\begin{tblr}{
  cells = {c},
  cell{3}{3} = {c=2}{},
  cell{5}{3} = {c=2}{},
  cell{6}{2} = {c=4}{},
  cell{7}{2} = {c=2}{},
  cell{7}{4} = {c=2}{},
  cell{8}{3} = {c=2}{},
  cell{9}{3} = {c=2}{},
  hlines,
  vlines,
  vline{2} = {-}{0.05em},
  hline{1-2,6,10} = {-}{0.08em},
}
 & $j \in \mu,\mathbb{A}$ & $j \in \rho,\mathbb{A}$ & $j \in \mu,\mathbb{B}$ & $j \in \rho,\mathbb{B}$  \\
$\qty[\hat M^z_{{\nu}}, \hat C^\pm_{{\mu,\rho}}]$ &     $ \pm \frac{\hbar}{2}   \hat C^\pm_{{\mu,\rho}}$   &   $ \mp \frac{\hbar}{2}   \hat C^\pm_{{\mu,\rho}}  $      &      $ \pm \frac{\hbar}{2}   \hat C^\pm_{{\mu,\rho}} $    & $ \mp \frac{\hbar}{2}   \hat C^\pm_{{\mu,\rho}} $\\
$\qty[\hat N^z_{{\nu}}, \hat C^\pm_{{\mu,\rho}}]$ &     $ \pm \frac{\hbar}{2}   \hat C^\pm_{{\mu,\rho}} $  &  $ \mp \frac{\hbar}{2}   \hat C^\pm_{{\mu,\rho}}   $  &    &   $     \pm \frac{\hbar}{2}   \hat C^\pm_{{\mu,\rho}}    $  \\
  $\qty[\hat N^\pm_{{\nu}}, \hat C^\pm_{{\mu,\rho}}]$ &           $0$  & $\pm \hbar \hat{N}^\pm_\mathrm{dia} \hat{S}_j^z$           &    $\mp \hbar \hat{N}^\pm_\mathrm{dia} \hat{S}_j^z$       &  $0$ \\
$\qty[\hat N^\pm_{{\nu}}, \hat C^\mp_{{\mu,\rho}}]$ &        $\pm \hbar \hat{N}^\pm_\mathrm{dia} \hat{S}_j^z$      &     $0$      &            &  $\mp \hbar \hat{N}^\pm_\mathrm{dia} \hat{S}_j^z$  \\
$\qty{\mathcal M^z_{{\nu}}, \mathcal C^\pm_{{\mu,\rho}}}_+$  &     $-\frac{\mathcal{C^\pm_{{\mu,\rho}}}\mathcal{S}^z_j}{2}$       \\
 $\qty{\mathcal N^z_{{\nu}}, \mathcal C^\pm_{{\mu,\rho}}}_+$ &      $-\frac{\mathcal{C^\pm_{{\mu,\rho}}}\mathcal{S}^z_j}{2}$       &    &     $+\frac{\mathcal{C^\pm_{{\mu,\rho}}}\mathcal{S}^z_j}{2}$   \\
  $\qty{\mathcal N^\pm_{{\nu}}, \mathcal C^\pm_{{\mu,\rho}}}_+$ &       $-\mathcal N^\pm_{{\nu}} \mathcal C^\pm_{{\mu,\rho}}$     &         $S(S+1)\hbar^2 \mathcal{N}^\pm_\mathrm{dia}-\mathcal N^\pm_{{\nu}} \mathcal C^\pm_{{\mu,\rho}}$ &            &      $-\mathcal N^\pm_{{\nu}} \mathcal C^\pm_{{\mu,\rho}}$        \\
$\qty{\mathcal N^\pm_{{\nu}}, \mathcal C^\mp_{{\mu,\rho}}}_+$ &          $S(S+1)\hbar^2 \mathcal{N}^\pm_\mathrm{dia}-\mathcal N^\pm_{{\nu}} \mathcal C^\mp_{{\mu,\rho}}$  &      $-\mathcal N^\pm_{{\nu}} \mathcal C^\mp_{{\mu,\rho}}$        &            &            $S(S+1)\hbar^2 \mathcal{N}^\pm_\mathrm{dia}-\mathcal N^\pm_{{\nu}} \mathcal C^\mp_{{\mu,\rho}}$
\end{tblr}
\caption{Commutation relations (dissipative brackets) between interlattice correlation operators (semiclassical correlations) and intralattice correlations. Commutators/brackets of intralattice correlations are not shown as only nearest-neighbor models are considered in the present paper. Here, $j$ denotes the shared spin of the two correlators at hand and "dia" refers to the diagonal bond connecting the two respective outer sites. The spin components $\hat{S}^z_j$ and $\mathcal{S}^z_j$ can be replaced by adjacent correlations using Eq.~\eqref{eq:replace-Sz}.}
    \label{tab:placeholder}
\end{table*}
\end{widetext}

%\nocite{*}

\bibliography{references}% Produces the bibliography via BibTeX.

\end{document}